\documentclass[fleqn,usenatbib]{mnras}

\usepackage{newtxtext,newtxmath}
\usepackage{makecell}
\usepackage{listings}
\usepackage{soul} 
\usepackage{cancel} 

\usepackage[T1]{fontenc}

\DeclareRobustCommand{\VAN}[3]{#2}
\let\VANthebibliography\thebibliography
\def\thebibliography{\DeclareRobustCommand{\VAN}[3]{##3}\VANthebibliography}

\newcommand{\aref}[1]{\hyperref[#1]{Appendix~\ref{#1}}}
\usepackage{graphicx}	
\usepackage{amsmath}	

\title[The Cluster Completeness Correction Calculator (C-4)]{The Cluster Completeness Correction Calculator (C-4): A Neural-Network framework and pilot application to the LEGUS Survey of NGC 628}

\author[Tang et al.]{Jianling Tang
$^{1}$\thanks{E-mail: janett.jianling@gmail.com}, Kathryn Grasha$^{1}$,
Tomasz Różański$^{1}$, Mark R. Krumholz$^{1}$
and Alan Zhang$^{1}$
\\
$^{1}$Research School of Astronomy and Astrophysics, Australian National University, Canberra, ACT 2611, Australia\\
}

\date{Accepted XXX. Received YYY; in original form ZZZ}

\pubyear{2026}

\begin{document}
\label{firstpage}
\pagerange{\pageref{firstpage}--\pageref{lastpage}}
\maketitle

\begin{abstract}
Integrated-light star cluster catalogues in external galaxies are subject to complex, often poorly-characterised selection effects that can bias inferred cluster demographics and introduce significant uncertainties, limiting the physical parameter space accessible to analysis. To mitigate this problem, here we introduce the Cluster Completeness Correction Calculator (C-4): a new software tool to quantify and predict these effects in both physical and photometric parameter spaces. C-4 adds artificial star clusters to observed galaxy images, processes these images through the same detection and filtering steps used to construct the original cluster catalogue, and then trains multilayer perceptron neural networks to learn the resulting selection function. The trained neural networks provide continuous, differentiable completeness functions that can be used for direct completeness corrections or incorporated into forward models. We present a pilot application of C-4 to NGC~628, demonstrating that the learned selection operator is highly accurate and successfully captures the strongly non-separable dependence of completeness on mass, age, and extinction. Applying the completeness correction to NGC 628 extends the range of cluster demographic analyses by roughly an order of magnitude in both mass and age, and removes artificial flattening in the observed cluster mass and age distributions. These results establish neural-network-based completeness modelling as a powerful and general approach for recovering intrinsic cluster populations, and provide a scalable framework for modelling high-dimensional selection functions in resolved stellar population studies.
\end{abstract}




\section{Introduction}
Star clusters are key laboratories for understanding how stars form and evolve within galaxies. Their masses, ages, and spatial distributions encode information about both local star formation efficiency and the large-scale assembly histories of their host galaxies. The demographic properties of star clusters, most prominently their mass and age distributions, provide quantitative constraints on cluster formation efficiency and disruption mechanisms, and how these processes vary with the larger galactic environment \citep[e.g.,][]{2019ARA&A..57..227K}. Robust demographic inference therefore depends critically on accurate modelling of survey selection effects -- yet in practice, this remains one of the dominant and least controlled sources of systematic uncertainty in cluster studies.

Over the past decade, high-resolution, multi-band imaging with HST and JWST imaging — including panchromatic imaging of M31 in PHAT \citep{PHAT}, UV-selected nearby galaxies in LEGUS \citep{2015LEGUS}, nearby spiral galaxies imaged with HST and JWST in PHANGS-HST and PHANGS-JWST \citep{PHANGS-HST, PHANGS-JWST}, and crowded-region imaging in PHATTER \citep{johnson_PHAT_2017} — have mapped star cluster populations in nearby galaxies with unprecedented detail. Despite these advances, observational catalogues remain incomplete: faint, extended, or highly extincted clusters are systematically underrepresented. Because demographic conclusions rely on subtle structure in mass and age distributions, even modest completeness biases can propagate into incorrect inferences about cluster disruption, environmental dependencies, and star formation physics. Accurately characterising completeness is therefore essential.  However, completeness is not a simple function of luminosity alone; it depends on cluster colour, size, light concentration, and the complex, spatially varying backgrounds against which clusters are detected \citep[e.g.,][]{2017ApJ...841...92R}.

Traditionally, completeness has been quantified using artificial star cluster tests (ASTs), in which synthetic clusters are injected into science images and recovery fractions are measured as a function of apparent magnitude \citep[e.g.,][]{2009Fall, 2010Chandar}. These curves are typically summarised as $50\%$ or $90\%$ detection limits and used to define nominally complete subsamples \citep[e.g., ][]{Fleming1995,Dolphin2000,Dalcanton2012}. While effective for simple truncation, this approach reduces a fundamentally high-dimensional selection process to a one-dimensional threshold, neglects correlations between cluster properties, and does not reproduce the full multi-band detection and catalogue construction pipelines used in modern surveys \citep[e.g,][]{adamo_legacy_2017, 2021Whitmore, 2025Hunt}.
Moreover, the logistic shape of the detection boundary implies that only a small fraction (often fewer than $\sim10\%$) of injected clusters probe the critical transition region, making AST sampling least informative precisely where accuracy is most needed.

A more fundamental limitation arises from the mismatch between where completeness is measured and where scientific inference is performed. AST-based estimates are defined in photometric space, whereas cluster demographics are inferred in the space of intrinsic physical parameters. This mismatch is not merely technical: it introduces an additional layer of marginalisation that is both computationally expensive and statistically unstable, particularly near the detection boundary where inference is most sensitive. The mapping from physical parameters to observable photometry is highly nonlinear and becomes intrinsically stochastic at low masses and young ages due to finite sampling of the stellar initial mass function (IMF; e.g., \citealt{Cervino03a, Cervino04a, 2012SLUGI, 2015SLUGIII}). As a result, clusters with identical mass and age can occupy broad regions of photometric space. If completeness is defined only in observables, demographic analyses must marginalise over this stochastic mapping, introducing additional computational cost and statistical instability, particularly near the sharp, logistic detection boundary where inference is most sensitive. In practice, finite AST sampling provides the least information precisely in this critical transition regime.

Recent work has begun to address some of these limitations. \citet{2024AJ....168...38H} introduced neural networks to learn photometric detection probabilities from ASTs, providing a flexible and continuous representation of multi-dimensional selection effects. However, these approaches model only photometric detection and do not learn the full catalogue selection operator, including subsequent filtering and quality-control classification steps. Consequently, they do not directly provide the quantity required for demographic inference: the probability that a cluster with given intrinsic physical properties is included in the final catalogue. 

To enable statistically rigorous cluster demographic inference, completeness must be expressed directly as a function of intrinsic cluster properties and must encapsulate the full catalogue construction process. 
In forward modelling and Bayesian hierarchical frameworks, one specifies a joint distribution of cluster mass, age, and extinction and evaluates the likelihood by mapping intrinsic parameters into observable photometry \citep[e.g.,][]{2012ApJ...750...60F, 2016Johnson}. If completeness is defined only in photometric space, this procedure requires an additional integration over the stochastic light-producing model, compounding computational costs and propagating systematic uncertainty into inferred cluster mass and age distributions. 

In this work, we overcome this limitation by developing a forward-modelling framework that learns the full catalogue selection operator, incorporating both photometric detection and all subsequent catalogue-level filtering steps. Rather than estimating completeness in photometric space and mapping it back to physical parameters, we directly learn the inclusion probability as a function of intrinsic cluster properties. This removes the need for additional marginalisation over stochastic light production and enables cluster-by-cluster completeness correction within a unified probabilistic framework. We embed our method in a publicly-released software package called \texttt{C-4}: The Cluster Completeness Correction Calculator.

We apply \texttt{C-4} to the LEGUS galaxy NGC 628 as a pilot case, for which completeness has previously been characterised using traditional methods, enabling a direct comparison. This provides a controlled benchmark demonstrating that our framework not only reproduces established results, but yields a smoother, more stable, and more physically consistent completeness model. We further present a first scientific application to correcting intrinsic cluster mass and age distributions using the published star cluster catalogue. Although demonstrated here for a single galaxy, the software is survey-agnostic and readily extensible to other datasets, including recently published JWST cluster catalogues \citep{2023Schinnerer, 2025Bortolini, maschmann2025ashescreationjwstuncovers, 2026Hassani} and future JWST surveys, where increased sensitivity and spatial resolution introduce additional challenges associated with stochastic sampling, crowding, and complex selection functions \citep{2015SLUGIII, adamo_legacy_2017}. We therefore anticipate that this method will provide a general tool for forward modelling of high-dimensional selection functions in extragalactic star cluster surveys.

The paper is structured as follows: in \autoref{sec:data}, we introduce the observed cluster catalogue used in this study. We describe the complete statistical framework and neural network architecture in \autoref{sec:pipeline}. Validation of method and corrected cluster mass and age distributions are presented in \autoref{sec:results}. We summarize the findings and discuss future prospects in \autoref{sec:conclusions}.

\section{DATA DESCRIPTION}
\label{sec:data}



LEGUS (Legacy Extragalactic UV Survey) is a Hubble Space Telescope Treasury program, providing imaging from the near–ultraviolet to the near-infrared with WFC3 (Wide-Field Camera 3) and ACS (Advanced
Camera for Surveys) in five broad-band filters (UV, U, V, R, and I).\footnote{Formally, these bands correspond to the F275W, F336W, F435W, F555W, and F814W filters on WFC3 and ACS. However, for brevity we will refer to these by the more conventional shorthands UV, U, V, R, and I.} The primary science goal of LEGUS is to characterise recent star formation and its connection to star clusters across diverse galactic environment in nearby galaxies. For a detailed description of the survey design, data reduction, and released products, we refer the reader to \citet{2015LEGUS}. 

In this pilot study we make use of the LEGUS data on NGC~628, a grand-spiral galaxy located at a distance of $\sim$9.8~Mpc \citep{2015LEGUS}. This galaxy has been the subject of extensive studies of its star formation properties, both from LEGUS and from numerous out studies. The LEGUS observation of this galaxy provides deep multi-band photometric observations from two pointings (C and E). Previous work using these data include several investigations of star cluster formation and evolution \citep{2015ApJ...815...93G,2017ApJ...840..113G,adamo_legacy_2017, Tang2023}. We adopt the central pointing (NGC~628C) as our pilot field for this work.

Because our objective is to learn the probability that a cluster enters the published catalogue for this field, it is essential to make explicit the full selection pipeline that defines catalogue inclusion. We therefore summarise the steps used to construct the catalogue here for reader convenience; full details are provided in \citet{2015LEGUS} and \citet{adamo_legacy_2017}. The LEGUS star cluster catalogue is constructed through a two-stage selection procedure, each stage introducing distinct observational and methodological selection effects. 

During the first stage, star cluster candidates are identified through automated detection and filtered using photometric and morphological cuts designed to remove obvious contaminants \citep[see for example][]{2015ApJ...815...93G, adamo_legacy_2017}. The steps carried out in this stage are as follows: 
\begin{enumerate}
\item Construct a white-light image, which is a weighted sum of the five filter images; the weighting scheme is described in \citet{2015LEGUS}, and we provide further details in \aref{app:white_light}.
\item Run Source Extractor \citep[\texttt{SExtractor};][]{1996SExtractor} on the white-light image to detect candidate clusters; the settings used in source extractor are described in \citet{adamo_legacy_2017}.
\item For each of the bounding apertures for candidates identified by Source Extractor, compute the photometric magnitude within that aperture on each of the five individual filter images, yielding magnitudes and uncertainties in each filter.
\item Discard candidates for which the computed photometric uncertainty in V-band is $> 0.3$ mag, or where the error in both B and I is $>0.3$ mag; objects discarded by this criterion are considered too faint to be identified as clusters with any confidence.
\item For the remaining candidates, compute the concentration index (CI) from the V-band image, defined as the magnitude difference between 1 and 3 pixel apertures. Discard those for which the CI is too high, indicating that the source is likely an individual star rather than a cluster; the CI cut value depends on the local background, and is determined following the procedure described in \citet{adamo_legacy_2017}.
\end{enumerate}
We refer to the catalogue of objects that remains after these two filtering steps as the ``automatic'' catalogue, since this step proceeds with no human intervention.


In the second stage, additional selections are applied to produce a candidate cluster catalogue, which consists of those objects that are considered bright enough to meaningful fits to their properties. From the automatic catalogue, in this stage we discard clusters that:
\begin{enumerate}
\item have photometric uncertainties $>0.3$ mag in two or more filters (implying that we retain only those with acceptably small uncertainties in at least four filters). This condition is imposed to obtain reliable constraints on the derived cluster properties (age, mass, extinction).
\item have absolute V-band magnitude $M_V > -6$. This cut does not represent the actual V-band detection limit; instead, it is a selection adopted by the LEGUS team to ensure that cluster candidates are sufficiently bright for reliable visual classification by humans. Since our goal is to reproduce this pipeline precisely, we make the same cut here, even though our procedure does not involve any human examination.
\end{enumerate}
We refer to the result after this stage as the cluster catalogue. This catalogue is then further subdivided into three classes based on a human analysis of their morphology; however, we will not attempt to replicate that human selection here, and will focus only on the steps leading up to construction of the cluster catalogue, not the sub-classes within it.

Each of the steps in catalogue construction imposes different and often hard-to-quantify biases and selection effects. For example, the CI cut likely selects against clusters that are too compact to distinguish from single stars at the available resolution. Similarly, the magnitude cuts clearly introduce a bias against both low-mass, old, and highly-extincted clusters. Taken together, the full pipeline therefore defines a high-dimensional selection operator that depends jointly on magnitude, colour, age, size, extinction, and local environment. In order to achieve our goal, our forward-modelling framework must learn all these biases.

\section{Completeness Measurement and Prediction Pipeline}
\label{sec:pipeline}

Here we introduce our method of measuring the completeness experimentally, and then building a neural network to predict the results of those experiments. We begin with a formal statement of the problem in \autoref{ssec:statement}, then describe our experiments in \autoref{ssec:asts} and the architecture of our neural network in \autoref{ssec:MLP_NN}.

\subsection{Formal statement of the problem}
\label{ssec:statement}

We first define the completeness function as the probability that a star cluster with a vector of properties $\boldsymbol{\theta}$ is included in the final LEGUS star cluster catalogue (\autoref{sec:data}). Specifically, we define a binary variable
\begin{equation}
C= 
\begin{cases}
1, & \text{cluster in the final LEGUS catalogue}, \\
0, & \text{otherwise},
\end{cases}
\label{eq:binary_inclusion}
\end{equation}
and define the catalogue-level completeness as
\begin{equation}
p_{\rm obs}(\boldsymbol{\theta}) \equiv P(C = 1 \mid \boldsymbol{\theta}),
\end{equation}
The vector of properties $\boldsymbol{\theta}$ can include both the intrinsic physical properties of the cluster -- mass, age, and extinction -- and the corresponding photometric magnitudes in the observed LEGUS filters.

For the reasons described in \autoref{sec:data}, this completeness function cannot be derived analytically. The intrinsic properties of the cluster population are unknown, and the full catalogue construction pipeline involves non-linear operations across multiple filters and selection criteria. As a result, the selection operator must be estimated empirically within a forward-modelling framework. The goal of the experimental framework we introduce in \autoref{ssec:asts} is therefore to measure the selection operator: for each test cluster with parameters $\boldsymbol{\theta}$, the pipeline applies the LEGUS selection criteria to return a binary catalogue-inclusion outcome $C \in \{0,1\}$. The resulting paired samples $\{(\boldsymbol{\theta}_i, C_i)\}$ therefore provide an empirical realisation of the mapping between cluster properties and catalogue inclusion, which the neural network we introduce in \autoref{ssec:MLP_NN} can then learn in order to predict the mapping
\begin{equation}
    \boldsymbol{\theta} \rightarrow P(C=1 \mid \boldsymbol{\theta}),
    \label{eq:prob_mapping}
\end{equation}
i.e., the probability that a cluster with a specified set of properties will be included in the catalogue.

The remaining choice to make is for which sets of intrinsic cluster properties to carry out this procedure. As noted above, for the $i^{\mathrm{th}}$ cluster the parameter vector is
\begin{equation}
\boldsymbol{\theta}_i = (\log M_i, \log T_i, A_{V,i}, \mathbf{m}_i),
\label{eq:theta_prop}
\end{equation}
where $M_i$, $T_i$, and $A_{V,i}$ denote the intrinsic mass, age, and extinction, respectively, and $\mathbf{m}_i = \{ m_{i,f} \}_{f \in \mathcal{F}}$ represents the synthetic photometric vector across the observed filter set $\mathcal{F}$. One could also add metallicity, or chemical abundances more broadly, as an additional intrinsic property, but for the purpose of this study we will assume that the metallicity is uniform across the population; we discuss the reasons for making this assumption, and its implications, in \autoref{ssec:metallicity}. In principle we could explore both the full set of elements of $\boldsymbol{\theta}$ and any subset thereof. In practice, however, not all combinations of elements are useful, and thus here we will focus on two cases: one where $\boldsymbol{\theta}$ includes only the physical parameters $(\log M_i, \log T_i, A_{V,i})$ and one where it includes only the photometric parameters $\mathbf{m}_i$. Each of these has a different use case: the mapping from physical parameters to completeness is the key quantity required if we wish to make completeness corrections on a distribution of clusters for which we have already estimated the mass, age, and extinction, while the mapping from photometric parameters to completeness is what we require in order to carry out forward modelling to compare a proposed theoretical cluster population to the observations. We will explore both cases in what follows.


\subsection{Artificial star cluster tests (ASTs)}
\label{ssec:asts}

To empirically sample the catalogue selection operator defined above, we perform artificial star cluster tests (ASTs) whereby we inject synthetic clusters spanning a broad range of physical and photometric properties into the observed data and then process the modified images through the same detection and filtering steps used to construct the LEGUS catalogue (\autoref{sec:data}). The resulting recovery outcomes provide labelled examples that map cluster parameters to catalogue inclusion. In this section we describe the steps required to carry out this test.

\subsubsection{Generation of artificial star clusters}
\label{ssec:generation_of_ASTs}

The first ingredient we require for ASTs is a set of artificial star clusters with known properties. For this purpose we generate a cluster library using the Stochastically Lighting Up Galaxies (\textsc{slug}) code \citep{2012SLUGI, 2015SLUGIII}. The code accepts as input a star cluster mass, age, and extinction, as well as a set of filters; it then operates by drawing a stellar population from the stellar initial mass function (IMF), using a set of stellar evolution tracks and stellar atmosphere libraries to calculate the spectrum emitted by each star (including contributions from nebular reprocessing), applying extinction, convolving the summed, extincted spectra with a specified set of photometric filters to generate predicted absolute photometry, then adding the NGC 628 distance modulus ($=29.98$ mag -- \citealt{2015LEGUS}) to convert absolute to apparent magnitudes. We refer readers to \citet{2015SLUGIII} for a full description of the pipeline, and document our choices for all \textsc{slug} parameters in \aref{app:slug_lib}. For the purpose of creating a set of clusters to use for artificial AST tests, we use \textsc{slug} to generate a library of $7.5\times 10^5$ synthetic clusters whose masses are uniformly sampled in logarithm from $10^2 - 10^8$ M$_\odot$, whose ages are uniformly sampled in logarithm from $10^5 - 1.5\times 10^{10}$ yr, and whose extinctions are uniformly sampled from $0 - 3$ mag. The stellar evolution tracks adopted in this work assume a solar metallicity of $Z=0.02$. This choice is consistent with the stellar population modelling assumptions used in previous studies \citep[e.g.,][]{2015SLUGIII, adamo_legacy_2017} and is appropriate for NGC 628, whose measured metallicity is approximately solar \citep{2013Berg}. Previous studies have shown that changing the assumed metallicity at the factor of two level has little effect on the inferred properties of the stellar population \citep{Krumholz15c}, but we defer a more detailed discussion of metallicity to \autoref{ssec:metallicity}.



In addition to these physical parameters used by \textsc{slug}, we require one more ``nuisance'' parameter: previous studies have shown that cluster effective radii affect the completeness function and subsequent star cluster population analyses \citep[e.g.,][]{2017ApJ...841...92R}. If cluster radii in turn are correlated with cluster masses or other properties, it is important to consider this effect when assessing completeness -- and there are good grounds to suspect that such a correlation is in fact present. Theoretically, cluster masses and radii should be correlated if clusters are tidally truncated by the galactic potential in which they reside. Observationally, heterogeneous cluster samples provide tentative evidence that star cluster masses are weakly correlated with cluster effective radii, with a large intrinsic scatter and a slope shallower than $r_{\rm eff} \propto M^{1/3}$ \citep{2019ARA&A..57..227K}.

To assess the consequences of such a correlation, in addition to a mass, age, and extinction, we also assign each cluster in our library an effective radius, and we consider two prescriptions for the mass–radius relation to adopt when assigning these effective radii. The first is a flat, mass-independent distribution of effective radii, and second is a Gaussian distribution in which the mean radius increases with mass as $r_{\rm eff} \propto M^{1/3}$. Specifically, for the first case we assume a uniform distribution from $r_\mathrm{eff} = 1 - 10$ pc, which for software implementation reasons we discuss in the next section we approximate as a discrete distribution 
\begin{equation}
    p(r_{\rm eff}) = \frac{1}{10}\sum_{i=1}^{10}\delta\!\left(r_{\rm eff}-i\,\mathrm{pc}\right),
    \label{eq:flat_mr}
\end{equation}
i.e., we choose the radius to be 1, 2, $\ldots$ 10 pc with equal probability. We choose this range to encompass the observed distribution of
effective radii measured for clusters in nearby galaxies which typically peak at a few parsecs and extend to $\sim10$ pc \citep{2017ApJ...841...92R}. For the second case, we carry out a very rough by-eye fit to the mass-radius relation shown in Figure 9 of \citet{2019ARA&A..57..227K}, and assign clusters radii
\begin{equation}
    \log_{10}(r_{\rm eff}/\mathrm{pc}) = 0.14\log_{10}(M/\mathrm{M}_\odot) + \varepsilon,
\label{eq:k2019_mr}
\end{equation}
where $\varepsilon$ is a random variable drawn from a Gaussian distribution with zero mean and a standard deviation of 0.21 dex; as with the first case, for reasons of ease of software processing we then round each cluster radius to the nearest whole number in units of pc from 1 to 10 pc. We do not attempt a more precise fit to the observed cluster mass-radius relation than this because the data available are highly heterogeneous in terms of both the location in which they are measured (Milky Way versus various external galaxies) and the analysis methods used to derive cluster radii; thus the data available do not justify a more precise fitting effort. Unless otherwise stated, we use the flat mass radius relation data set for the results presented below. However, we also present a comparison of the results to assess how sensitive they are to the assumed mass-radius relation in \aref{app:mass-radius}.

\subsubsection{Generation of synthetic images}

The second step in our AST pipeline is to insert synthetic clusters into images to generate modified images suitable for retrieval testing. The process to generate a single set of test frames is as follows. We first select a value of $r_\mathrm{eff}$ and then randomly select (without repeats) 500 clusters from the subset of our library with that value of $r_\mathrm{eff}$; we choose 500 because we wish the number to be as large as possible for reasons of computational efficiency (since the cost of cluster insertion and finding is roughly fixed per-frame), but we need to keep the number of clusters inserted per test small enough to ensure that our artificial clusters are greatly outnumbered by real ones, and thus do not meaningfully increase the level of crowding. The LEGUS catalogue for NGC 628 contains $\sim 1300$ clusters, so 500 represents a reasonable compromise between these priorities.

We next assign positions within the images to these clusters. It is important that we not simply assign these positions purely randomly, because local crowding is a key factor affecting completeness -- it is more difficult to recover clusters located in regions with high background stellar light contributed by neighbouring sources than those in regions of low local background. To ensure that the distribution of local crowding is realistic, we convolve the white light image of the target field with a Gaussian kernel of width $\sigma = 120$ pc to construct a smoothed representation, then draw cluster positions in the image with a probability density proportional to the smoothed light intensity. The choice of 120 pc as the smoothing scale is again a compromise, and is motivated by both empirical and physical considerations. Empirically, using values substantially smaller than this results in a majority of our synthetic clusters being placed exactly on top of real ones, while using a value that is much larger washes out the structure in the galaxy so much that we are effectively placing the clusters near-randomly; either would defeat the realism of the test. Physically, 120 pc is about the scale height of the star-forming phase of the interstellar medium \citep[e.g.,][]{Boulares90a}, and thus is about the characteristic size scale we expect for star-forming molecular complexes; our choice therefore smooths over the internal structure of these complexes while preserving their presence. Finally, in order to avoid overlaps between our artificial clusters, we additionally enforce a minimum spacing of three times the cluster effective radius; thus if when we draw a position for one of our synthetic clusters the result is within three effective radii of a position we have already drawn, we reject it and draw a new position.

Once we have chosen positions within the image, next step is to generate the light distribution describing individual clusters. Following the approach previously used in LEGUS \citep{2017ApJ...841...92R}, we assume that the intrinsic light profile of clusters follows an \citet[hereafter EFF]{EFF1987} distribution, which has a flat central core and a powerlaw envelope characterised by two parameters: a core radius and a powerlaw slope that describes the shape of the profile outside the core region. We adopt $\eta = 1.5$ for the envelope slope, and fix the core radius by setting the half-light radius of the cluster equal to $r_\mathrm{eff}$. We choose the EFF profile because extragalactic clusters younger then $\sim 0.5$ Gyr have not yet lost their envelopes to tidal stripping, and thus tend to have powerlaw-like light profiles rather than the more sharply truncated ones that characterise older clusters \citep{2019ARA&A..57..227K}; for this reason, EFF-type profiles have been widely used in \textit{HST}-based size measurements and ASTs for such systems.

We then convolve this intrinsic profile with an empirical \textit{HST} stellar PSF describing the \textit{HST} resolution to produce combined profiles for the light in each filter image. The empirical PSF for each filter is supplied to \texttt{BAOLab} \citep{Larsen14a} as a user-defined PSF and is constructed from 20 visually selected isolated stars identified in the corresponding \textit{HST} science images using the \texttt{IRAF}'s \texttt{daophot} \citep{1986IRAFI} PSF-fitting routine. This approach ensures that the adopted PSF captures the actual observational characteristics of the science images, including the telescope optics, detector response, and image-processing effects associated with the final reduced data products. We then carry out this convolution using the \texttt{mkcmppsf} routine in the \texttt{BAOLab} image processing software.   The resulting artificial clusters have the same pixel grid as the science images. For NGC~628C, the science frames have dimensions of $7200\times7200$ pixels whereas the adopted pixel scale is $0.04'' \mathrm{pixel}^{-1}$ for WFC3 images and $0.05'' \mathrm{pixel}^{-1}$ for ACS images. 

Finally, we again use \texttt{BAOLab} to insert the synthetic clusters into each image filter image at the chosen set of positions and magnitudes in each filter, via its \texttt{mksynth} routine. The intrinsic cluster profiles are converted from physical units (pc) to image-plane scales using the galaxy distance and the instrument pixel scales before being injected into the science images. This step properly accounts for the sensitivity of the observation by directly sampling from the cluster light profile with the appropriate level of shot noise. Note that, due to the limitations of \texttt{BAOLab} we are required to use the same light profile for every synthetic cluster, which is why we batch the clusters into ten bins of fixed $r_\mathrm{eff}$. At this point we have a set of science-ready images with synthetic clusters added.

\begin{figure}
    \centering
    \includegraphics[width=\linewidth,height=0.8\textheight,keepaspectratio]{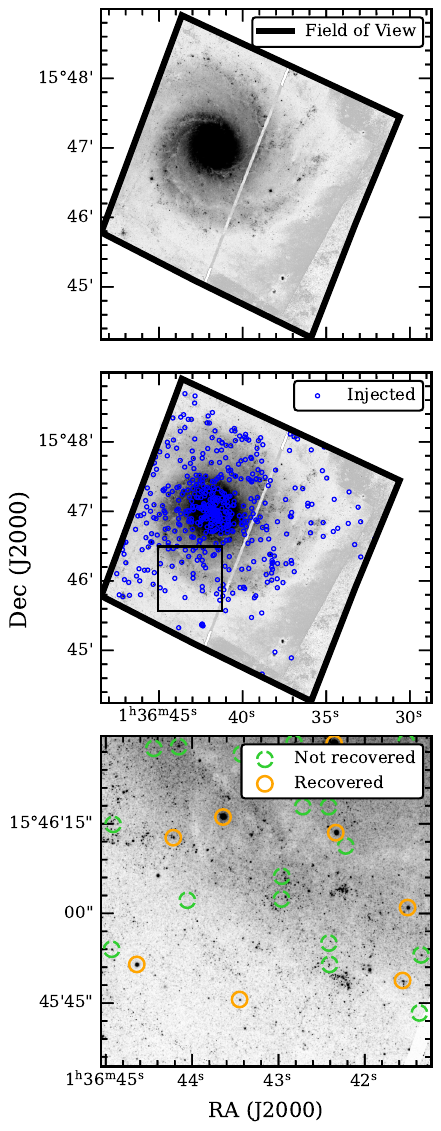}
    \caption{Illustration of the AST pipeline for artificial clusters with $r_{\rm eff}=5\,\mathrm{pc}$.
    \emph{Top:} white-light image of NGC~628 from the original LEGUS survey, with the ``C'' (central) field of view that we use in this paper outlined in black. The white streak across the image is an instrumental artefact caused by the chip gap in the ACS image, while the overlapping border in the lower-right corner is a stacking artefact arising from the combination of ACS and WFC images with slightly different pointings.
    \emph{Middle:} same image with 500 injected synthetic clusters overlaid as blue circles. The black box marks the zoomed-in region shown in the bottom panel.
    \emph{Bottom:} zoom-in showing the catalogue-level selection outcome, with recovered objects marked by orange circles and non-recovered objects by green dashed circles.
    All panels share a common WCS projection in RA and Dec (J2000).}
    \label{fig:synthetic_clusters}
\end{figure}

We show an example of this procedure in \autoref{fig:synthetic_clusters} for a representative frame containing 500 synthetic clusters with an effective radius of 5 pc. The $r_{\rm{eff}}$ value of 5 pc is shown for illustrative purposes only; the full AST suite spans effective radii from 1 to 10 pc as described in \autoref{ssec:generation_of_ASTs}. The top panel shows the white-light image of NGC 628C, with the black outline marking the field of view. The middle panel shows the positions of the 500 injected synthetic clusters overlaid on the same image. As expected, since these positions are drawn from a probability distribution based on the smoothed white-light profile, more clusters are placed in crowded, high-surface-brightness regions to mimic the underlying crowding conditions of the galaxy.

\subsubsection{Processing of synthetic images}

The final step in our pipeline is to process our synthetic images through the same detection and filtering steps used in the construction of the LEGUS cluster catalogue. Our procedure for this is identical to that used in the original LEGUS processing pipeline, involving the same five steps outlined in \autoref{sec:data} -- constructing a white-light image, running \texttt{SExtractor} to find sources on the white light image, performing aperture photometry on the individual filter images at the coordinates of recovered cluster candidates, discarding candidates for which the photometry is too uncertain, and discarding those for which the concentration index is too large. We have verified that, if we run our software pipeline on unmodified LEGUS science frames (i.e., frames for which we have not added artificial clusters), the outputs at each step match the archived LEGUS results. In particular, our white light images are identical to the ones available from the LEGUS website, and our final cluster catalogue matches the official LEGUS catalogue.

Once we have completed all these steps on a set of test frames, we check which of our synthetic clusters have been recovered. We consider a cluster to have been recovered if the final catalogue produced at the end of our procedure includes a cluster that is within three pixel of the coordinates at which we inserted an artificial cluster. This yields a binary inclusion label $C$ for each injected synthetic cluster, corresponding to the catalogue-level selection variable defined in \autoref{eq:binary_inclusion}. Thus the final output of this step is a set of 500 synthetic clusters with known physical and photometric properties labelled with $C=0$ or $C=1$.  The bottom panel of \autoref{fig:synthetic_clusters} shows a zoomed-in region highlighting the final output of the completeness pipeline, with recovered clusters $C=1$ marked by orange circles and non-recovered clusters $C=0$ by blue dashed circles. The recovered clusters generally coincide with visually identifiable sources, providing a qualitative validation of the extraction procedure.

\label{ssection:LEGUS_pipeline}

\subsubsection{Data set generation}
\label{ssec:data_generation}

To produce our main data set for the purposes of training the neural network, we use the procedure outlined above to generate 49 sets of test frames for each of our 10 possible values of $r_\mathrm{eff}$; since each set of test frames contains 500 clusters, this means that our main training set consists of 245,000 clusters. We ensure that selection of artificial clusters is done with no repeats not only within a single set of test frames but between frames, i.e., each of the 245,000 clusters represents a different draw from the library.

We also produce a secondary test data set that we will use to validate the neural network. For the test data we generate 100 sets of test frames for each possible $r_\mathrm{eff}$, but we use the \textit{same} set of 500 clusters for each of the 100 test sets. Thus we have only 5000 distinct clusters (500 per frame times 10 values of $r_\mathrm{eff}$), but with each cluster repeated 100 times, using new random positions and random realisations of the shot noise during each repeat. This allows us to measure $p_\mathrm{obs}$ \textit{directly} for each of our 5000 test clusters, simply by checking in what fraction of the 100 trials it was recovered. We ensure that there is no overlap between clusters in the test and training sets, i.e., if we pick a particular cluster to use in the training set then it is ineligible to be one of the 5000 in our test set.

\subsection{Modelling the completeness function using multi-layer perceptron neural networks}
\label{ssec:MLP_NN}

\subsubsection{MLP problem formulation and architecture}

The artificial star cluster tests described above produce 245,000 labelled pairs $(\boldsymbol{\theta}_i, C_i)$, where $\boldsymbol{\theta}_i$ denotes the cluster properties and $C_i \in \{0,1\}$ indicates catalogue inclusion. (We set aside for the moment the 5000 clusters that we have repeated 100 times; we make use of them only for testing, not for network training or validation.) 

Formally, we regard $C\in\{0,1\}$ as a Bernoulli random variable whose only stochastic degrees of freedom in our AST pipeline are the injection position $s_1$ and realisation of shot noise $s_2$. These are drawn from distributions encoding, respectively, the local background and crowding and the image shot noise. For notational simplicity, we denote the joint distribution of these two random variables $P(s_1, s_2)$, though we note that they are drawn independently and thus their joint PDF could be written as a product of individual marginal PDFs. Let $\mathcal{S}(\boldsymbol{\theta}, s_1, s_2)\in\{0,1\}$ denote the deterministic output of the LEGUS selection pipeline for a cluster with parameters $\boldsymbol{\theta}$, injected at position $s_1$ and with shot noise realisation $s_2$. Here $s_2$ denotes a particular realisation of the Poisson photon-counting noise introduced during artificial cluster image generation. In our AST pipeline, this noise is generated when \texttt{BAOLab} inserts synthetic clusters into the science images by drawing Poisson realisations of the expected pixel-level photon counts derived from the cluster light profile. Consequently, otherwise identical clusters may yield different observed fluxes and catalogue-inclusion outcomes due solely to different noise realisations. We define completeness as the inclusion probability marginalised over the random cluster placement and shot noise,
\begin{equation}
p_{\mathrm{obs}}(\boldsymbol{\theta})
\equiv
\mathbb{P}(C=1\mid \boldsymbol{\theta})
=
\mathbb{E}_{s_1,s_2\sim P(s_1,s_2)}
\!\left[
\mathcal{S}(\boldsymbol{\theta}, s_1, s_2)
\right].
\label{eq:pobs_def}
\end{equation}
where $\mathbb{E}_{s_1, s_2\sim P(s_1,s_2)}[\cdot]$ denotes expectation values of inclusion probabilities of clusters with parameters $\boldsymbol{\theta}$ with respect to the distribution $P(s_1,s_2)$ of position and shot noise.

Given this mathematical description, completeness estimation can be cast as a supervised binary classification problem. We model this mapping using a multi-layer perceptron \citep[MLP;][]{rumelhart1986backprop} implemented in PyTorch\footnote{Pytorch: \url{https://pytorch.org/}} \citep{paszke2019pytorch}. The input parameter space in our problem is low-dimensional, consisting either of a small set of physical parameters $\boldsymbol{\theta}_{\rm phys} = (\log M, \log T, A_V)$ or a photometric vector $\boldsymbol{\theta}_{\rm phot} = \mathbf{m}$. In this regime, MLPs provide a simple and well-understood function approximator capable of representing highly non-linear decision boundaries while remaining computationally efficient. Importantly, their representational capacity can be controlled through a small number of hyperparameters, primarily the number of layers and neurons, as well as regularisation strength, allowing the model complexity to be matched to the smoothness of the underlying completeness function without introducing unnecessary architectural overhead.

\begin{figure}
    \centering
    \includegraphics[width=0.53\textwidth]{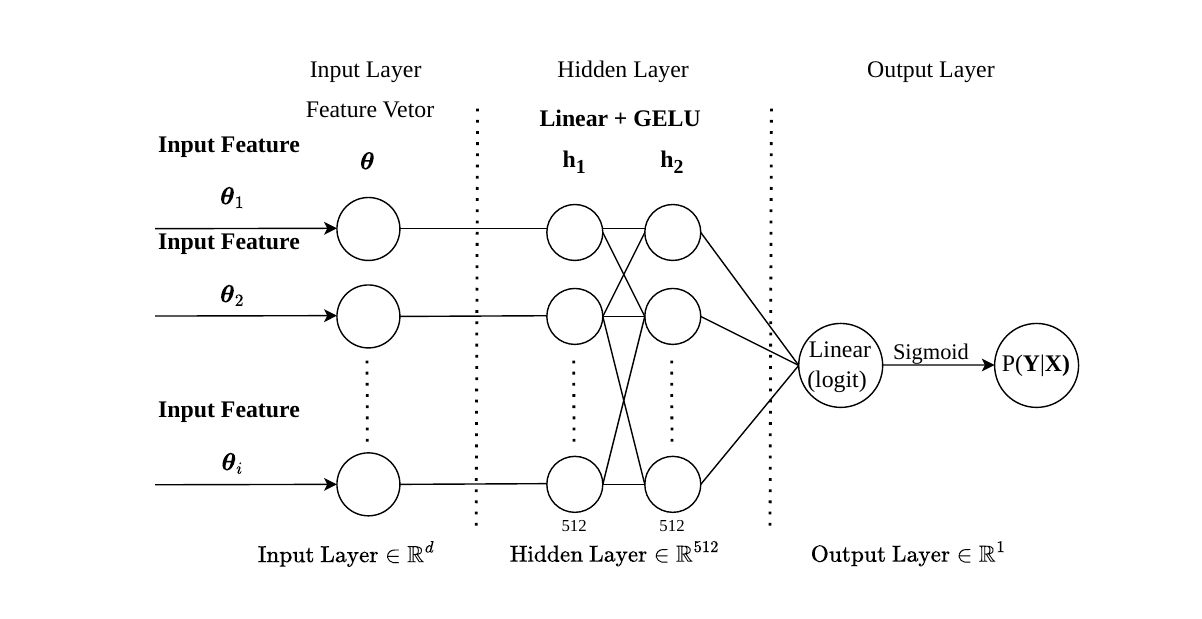}
    \caption{An illustration of neural network used to estimate completeness. The network takes as input a parameter vector $\theta \in \mathbb{R}^d$, where $d$ denotes the dimensionality of the input feature space ($d=5$ for photometric input vectors and $d=3$ for physical input vectors). The input vector is passed through two fully connected hidden layers, each containing 512 neurons and employing GELU activation functions. A final linear layer followed by a sigmoid activation outputs the predicted completeness probability.}
    \label{fig:NN_architecture}
\end{figure}

We train two distinct neural networks to accept the two sets of parameters, and refer to them from this point forward as the ``physical'' and ``photometric'' networks. For both networks we define
\begin{equation}
{\rm MLP}_{\boldsymbol{\phi}} : \boldsymbol{\theta} \mapsto \hat{p}_{\rm obs} \in [0,1],
\end{equation}
where $\boldsymbol{\phi}$ denotes the free parameters of the MLP trained to approximate the completeness function. Mathematically, the MLP for this binary classification task is implemented as a composition of several affine transformations, followed by element-wise nonlinear activations, with a logistic function applied to the final scalar output (logit $z$) transforming it into estimated probability. In our case, we use an MLP with two hidden layers, each containing 512 neurons, and GELU nonlinearity \citep{hendrycks2016gelu}. Writing the equations explicitly for this architecture, given an input vector $\boldsymbol{\theta}$ (either $\boldsymbol{\theta}_{\rm phys}$ or $\boldsymbol{\theta}_{\rm phot}$), the network computes
\begin{equation}
\begin{aligned}
\mathbf{h}_1 &= \mathrm{GELU}\!\left( \mathbf{W}_1 \boldsymbol{\theta} + \mathbf{b}_1 \right), \\
\mathbf{h}_2 &= \mathrm{GELU}\!\left( \mathbf{W}_2 \mathbf{h}_1 + \mathbf{b}_2 \right), \\
z &= \mathbf{w}_3^\top \mathbf{h}_2 + b_3, \\
\hat{p}_{\rm obs} &= \sigma(z),
\end{aligned}
\label{eq:NN}
\end{equation}
where $\mathbf{W}_1$ and $\mathbf{W}_2$ denote the trainable weight matrices, $\mathbf{b}_1$ and $\mathbf{b}_2$ the corresponding bias vectors, $\mathbf{w}_3$ the output-layer weight vector, and $b_3$ the scalar output bias. The scalar logit $z$ is transformed into the estimated probability $\hat{p}_{\rm obs}$ using the logistic function, $\sigma(x) = 1 / (1 + e^{-x})$. A schematic view of the adopted MLP architecture is shown in \autoref{fig:NN_architecture}.

\subsubsection{Training and model selection}

We split the primary dataset, composed of $2.45 \times 10^5$ pairs $(\boldsymbol{\theta}_i, C_i)$, into 80\% / 20\%  train / validation subsets using a stratified shuffle split to preserve class balance \citep{pedregosa2011scikit}. In both parameterisations, input vectors are standardised using the mean and variance computed from the training set \citep{goodfellow2016deep}.

After splitting the data set, we optimise network parameters by minimising the binary cross-entropy loss. Although the network is written in \autoref{eq:NN} as producing the estimated inclusion probability $\hat{p}_{\rm obs} = \sigma(z)$, in practice the loss is computed directly from the corresponding logits $z$ using \texttt{BCEWithLogitsLoss} in PyTorch \citep{paszke2019pytorch}, which combines the logistic transformation and binary cross-entropy in a numerically stable form.

Optimisation is performed using the AdamW optimiser \citep{loshchilov2019adamw}, a first-order gradient-based optimisation method. Model parameters are updated iteratively using  gradients estimated from mini-batches of the training data. We adopt a learning-rate schedule consisting of a linear warm-up over the first 10\% of training steps followed by cosine decay \citep{loshchilov2017sgdr,vaswani2017attention}. We provide full details on our network training approach, including hyperparameter optimisation and loss curves, in \autoref{app:training}.

\section{Results}
\label{sec:results}

Here we evaluate the performance of the neural-network completeness model and demonstrate its application to cluster demographics. In \autoref{ssec:results_validation} we validate the network predictions against independent test data. In \autoref{ssec:results_surface} we examine the inferred completeness as a continuous function in physical parameter space. Finally, in \autoref{ssec:results_demographics} we apply the model to derive completeness-corrected cluster age and mass distributions.

\subsection{Validation of the neural-network}\label{ssec:results_validation}

Before using the neural networks as a selection function in completeness correction, we test how well they perform by comparing their predictions to our test set of 5000 clusters repeated 100 times each (\autoref{ssec:data_generation}). As a first step in this testing, we use both our physical and photometric networks to predict the value of $p_\mathrm{obs}$ for each test cluster; we refer to these predictions as $\widehat{p}_\mathrm{phys}$ and $\widehat{p}_\mathrm{phot}$, respectively. In the tests that follow, we will compare these predictions to the true, measured completeness, which for each test cluster we define as
\begin{equation}
    p_\mathrm{true} = \frac{1}{N_\mathrm{rep}} \sum_{n=1}^{N_\mathrm{rep}} C_n,
\end{equation}
where $N_\mathrm{rep} = 100$ is the number of repetitions and $C_n$ is the value of $C$ ($=0$ for a non-detection and $=1$ for a detection) in the $n$th trial.

\subsubsection{Binned comparisons}

\begin{figure}
    \centering
    \includegraphics[width=0.5\textwidth]{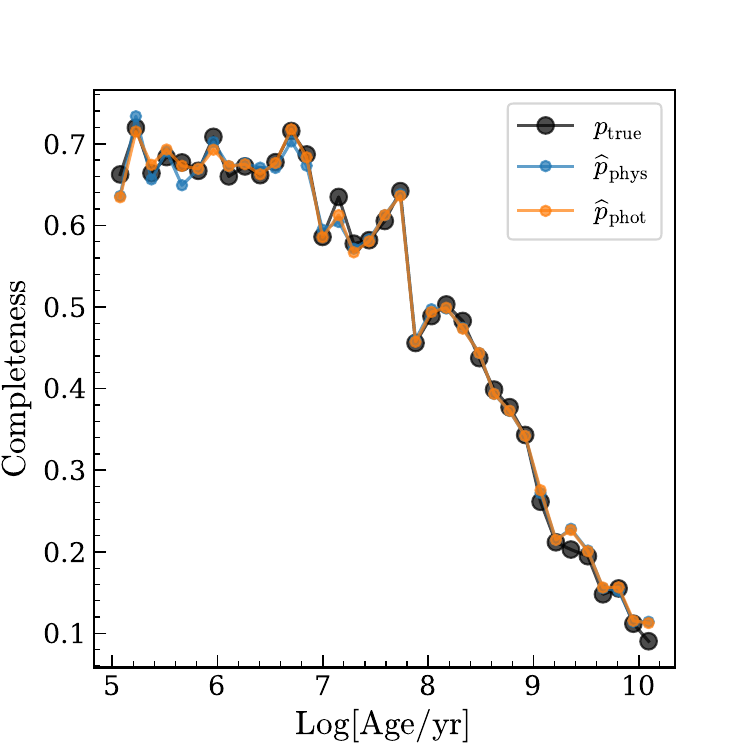}
    \caption{Arithmetic means of the true completeness, $p_\mathrm{true}$ (black points), and the completeness values predicted by the physical and photometric neural networks, $\widehat{p}_{\mathrm{phys}}$ (blue points) and $\widehat{p}_\mathrm{phot}$ (orange points), for the test cluster sample binned by log cluster age.}
    \label{fig:validation_agebin}
\end{figure}

\begin{figure}
    \centering
    \includegraphics[width=0.5\textwidth]{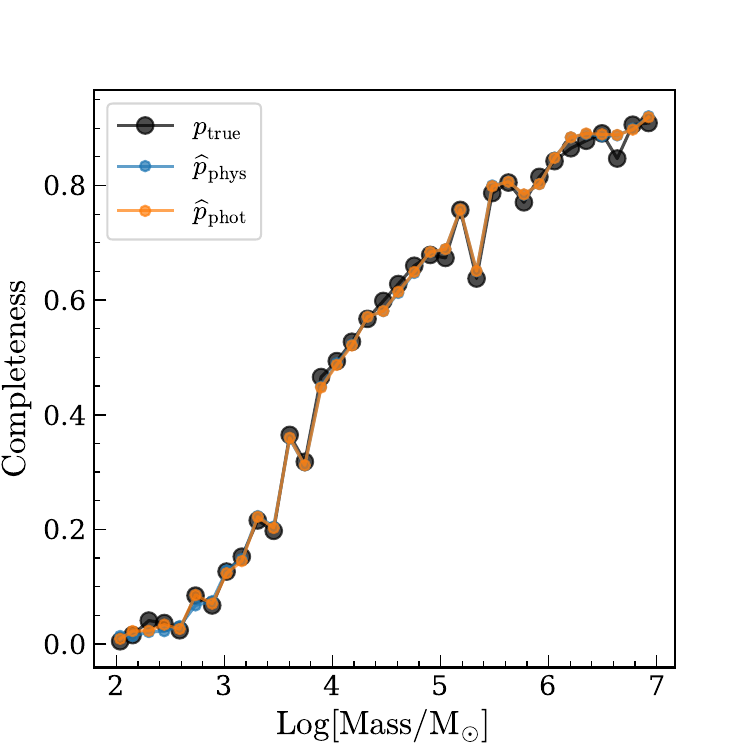}
    \caption{Same as \autoref{fig:validation_agebin}, but binned in $\log M$ rather than log age.}
    \label{fig:validation_massbin}
\end{figure}

\begin{figure}
    \centering
    \includegraphics[width=0.5\textwidth]{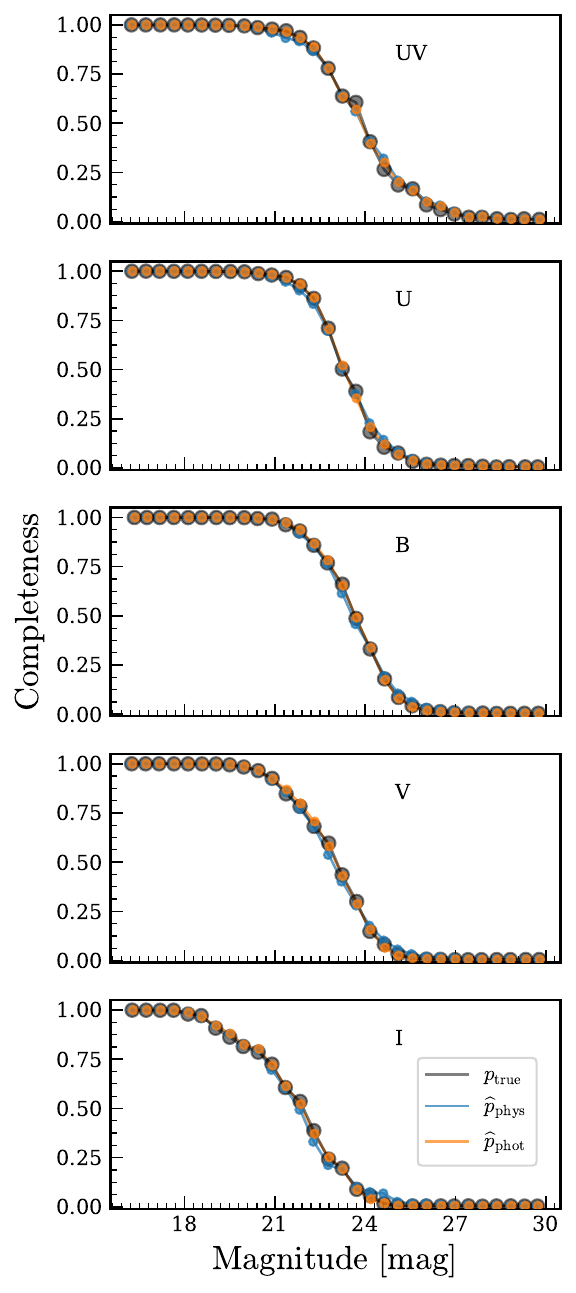}
    \caption{Same as \autoref{fig:validation_agebin}, but binned by magnitude in the UV (F275W), U (F336W), B (F435W), V (F555W), and I (F814W) filters, from top to bottom.}
    \label{fig:validation_magnitudes}
\end{figure}

For our first comparison we divide the test data set into uniformly-spaced bins by one variable of interest -- log mass, log age, or magnitude in one of the available photometric filters -- and compare the arithmetic mean value of $p_\mathrm{true}$ for clusters in that bin to the mean of the neural network-predicted values $\widehat{p}_\mathrm{phys}$ and $\widehat{p}_\mathrm{phot}$. For log mass and log age we use 35 uniformly-spaced bins, while for photometric magnitude we use 60 uniformly-spaced bins. The test set spans the full physical parameter space, covering cluster masses from $10^{2}$ to $10^{7}\,{\rm M}_\odot$ and ages from $10^{5}$ to $10^{10}\,{\rm yr}$, and the full range of photometric magnitudes from $\approx 16 - 30$, ensuring that the evaluation is not restricted to a limited region of parameter space. We omit bins containing fewer than 80 clusters from our plots to ensure that we are not biased by small $N$ statistics. We show the results for bins in age (\autoref{fig:validation_agebin}), mass (\autoref{fig:validation_massbin}), and photometric magnitude across all the five available filters (\autoref{fig:validation_magnitudes}). In each case, black points denote the mean $p_\mathrm{true}$, while blue and orange points show the corresponding means of $\widehat{p}_\mathrm{phys}$ and $\widehat{p}_\mathrm{phot}$.

Overall, both the physical and photometric networks show good agreement with the true completeness across the full range of masses, ages, and magnitudes. The networks recover not only the mean trend, but also sharp features -- for example a sharp drop in completeness at $\approx 10^{7.6}\,{\rm yr}$ in the age-binned comparison (\autoref{fig:validation_agebin}), and the sawtooth structure at intermediate F814W magnitude (bottom panel of \autoref{fig:validation_magnitudes}). Notably, the physical neural network achieves performance comparable to the photometric network despite relying only on physical parameters without direct access to photometric observables -- which are the quantities that more directly determine whether a cluster is recoverable or not. 
This indicates that the model successfully captures the underlying mapping between physical and photometric representations of cluster properties, and
preserves sufficient structure to define a well-constrained decision boundary in completeness space without direct access to photometry. Indeed, the only noticeable systematic difference between the predictions of the physical and photometric networks occurs for young and low-mass clusters, where 
incomplete sampling of the IMF introduces non-deterministic mappings between cluster physical and photometric properties, and thus this mapping is hardest to learn -- but even in this regime the differences are small. We explore the performance of our neural network in the this regime further in \autoref{ssec:quantitative_comparison}.



\subsubsection{Cluster-by-cluster comparison}
\label{ssec:quantitative_comparison}
\begin{figure}
    \centering
    \includegraphics[width=0.5\textwidth]{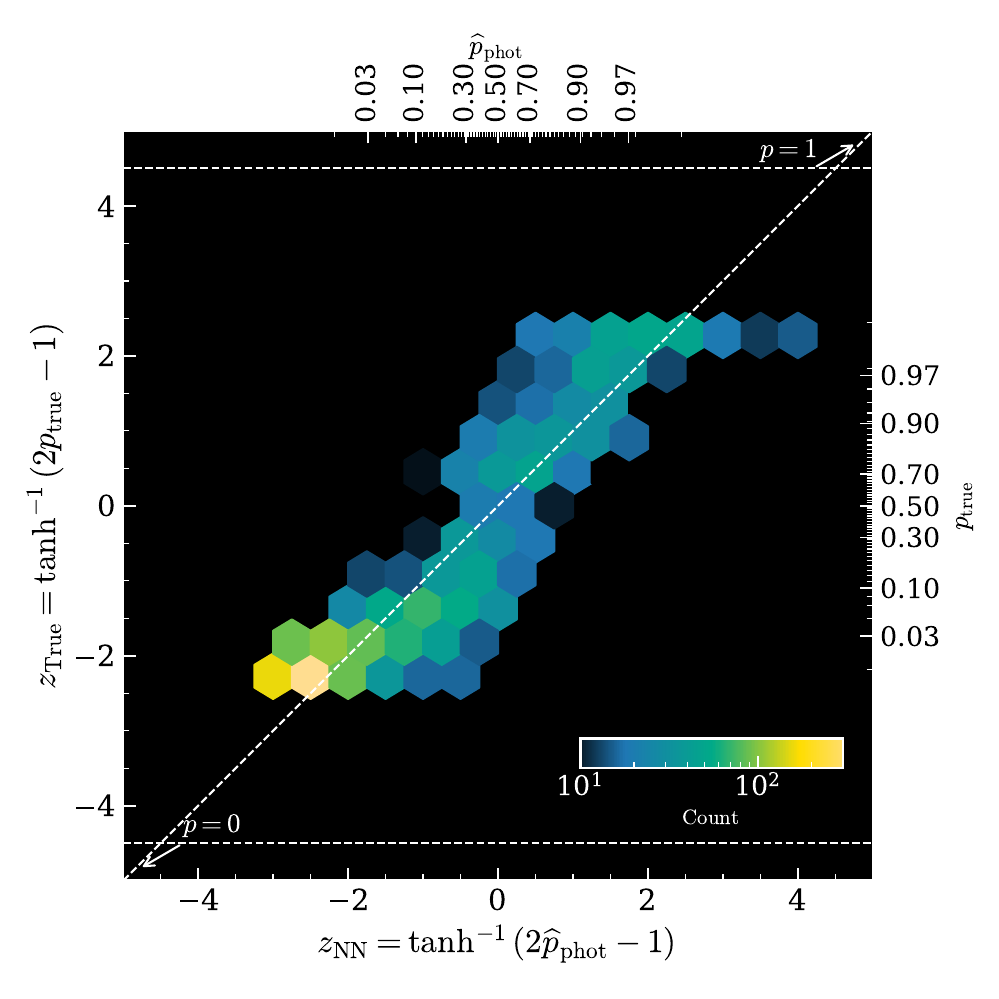}
    \caption{Counts of test clusters in hexagonal bins (see colour bar) in the space of neural network-predicted completeness (horizontal axis) versus true completeness (vertical axis). Rather than plot the true completenesses $p$, we plot positions in the transformed variables $z$ (given by \autoref{eq:atanh_transform}) to avoid crowding near probabilities of zero and unity. The value of true probability $p$ corresponding to the coordinates $z$ are indicated on the top and right axes. The white dashed line shows the one-to-one relation and colours indicate log-scaled counts in each bin. Note that clusters for which the observed or true or predicted completeness is exactly 0 or 1 are omitted from the plot, since these values are mapped to $z=\pm\infty$. Two additional horizontal white dashed lines, together with white arrows, indicate the directions corresponding to completeness of 0 and 1.}
    \label{fig:validation_1to1_hexbin}
\end{figure}

The binned comparisons in the previous section show that our neural networks do an excellent job of reproducing the statistical average completeness for slices of the cluster population, which is sufficient for most analysis purposes. However, it is also helpful to characterise the typical level of error in prediction for a single cluster, rather than the mean of a population. To this end, in addition to the binned comparisons in physical and photometric space, we also test our neural networks by comparing the predicted and observed completenesses cluster-by-cluster. As a first step toward this comparison, we can directly plot $p_\mathrm{true}$ versus $\widehat{p}_\mathrm{phot}$, since the photometric neural network achieves slightly better predictive performance than the physical neural network. Visualising this comparison is challenging due to the large number of points for which both the predicted and observed completeness are very close to zero or to unity, so to mitigate this we apply the transformation
\begin{equation}
z \equiv \tanh^{-1}(2p-1),
\label{eq:atanh_transform}
\end{equation}
where $p$ is the true or predicted completeness. This transformation maps $p=0.5$ to $z=0$ but $p = 0$ or 1 to $-\infty$ and $+\infty$, and thus stretches out the distribution near the boundaries, allowing a clearer view. We show the true versus predicted completenesses transformed in this manner in \autoref{fig:validation_1to1_hexbin}, using a heatmap in hexagonal bins to show the shape of the distribution; we suppress bins containing fewer than 30 clusters, ensuring that the density structure is not dominated by Poisson fluctuations in low-count cells. 
We see that the distribution exhibits no significant systematic bias: the highest-density bins lie along the one-to-one relation, and the residual scatter appears approximately symmetric about this line across the dynamic range. This behaviour demonstrates that the photometric NN accurately reproduces cluster-level completeness without introducing systematic bias in probability space. 
 
\begin{figure}
    \centering
    \includegraphics[width=1\linewidth]{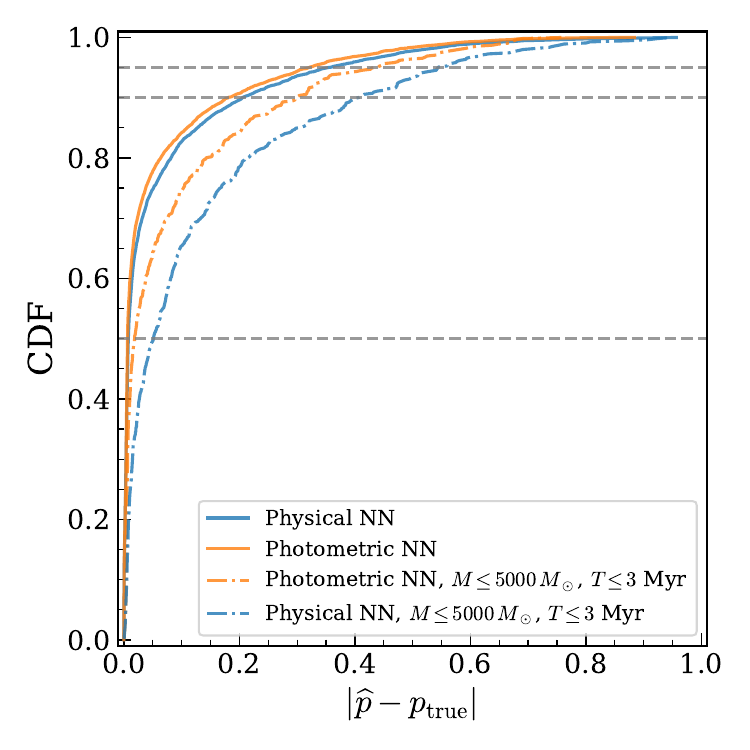}
    \caption{Cumulative Distribution Function (CDF) of the absolute prediction error $\left|\widehat{p}_\mathrm{\{phys,phot\}} -p_{\mathrm{true}} \right|$ for the physical (blue) and photometric (orange) neural networks evaluated on the test set. Solid lines indicate results for the full test est, while dot-dashed lines show results for the low-mass ($M\leq 5000$ M$_\odot$), young-age ($T \leq 3$ Myr) part of the test set only, where stochastic effects are expected to be strongest. The dashed horizontal lines mark the $50^{\mathrm{th}}$, $90^{\mathrm{th}}$, and $95^{\mathrm{th}}$ percentiles of the error distribution for the full test set (solid lines), from which the corresponding error thresholds can be read directly from the horizontal axis.}
    \label{fig:CDF_error}
\end{figure}

To quantify the scatter around the one-to-one line, we examine the cumulative distribution function (CDF) of the absolute prediction error, defined as
\begin{equation}
\Delta p_\mathrm{\{phys,phot\}} \equiv \left| \widehat{p}_\mathrm{\{phys,phot\}} - p_{\mathrm{true}} \right|.
\label{eq:delta_p}
\end{equation}
For any threshold $\delta$, the CDF gives the fraction of clusters with $\Delta p_\mathrm{\{phys,phot\}} \leq \delta$. The CDFs (\autoref{fig:CDF_error}) demonstrates that prediction errors are minimal for the majority of clusters: for the physical network, the $50^{\mathrm{th}}$, $90^{\mathrm{th}}$, and $95^{\rm{th}}$ percentiles of the absolute error distribution $\Delta p$ are 0.007, 0.200, and 0.356, respectively; for the photometric neural network, they are 0.006, 0.177, and 0.321. The corresponding curves show that the photometric neural network yields a modestly tighter error distribution overall, consistent with its direct access to the observables that govern catalogue inclusion, but both models provide reliable cluster-level completeness estimates. This indicates that the network is well-calibrated in probability space.

We can use the same approach to examine in more detail the performance of our neural networks in the regime where stochastic effects are expected to be largest, for young, low-mass clusters, defined as $\mathrm{age}<3\times 10^6\,\mathrm{yr}$ and $M<10^{3.5}\,M_\odot$. There are $N=551$ such clusters in our test sample, and we plot the CDFs of physical and photometric neural network errors for this subset as dot-dashed lines in \autoref{fig:CDF_error}. We see that the errors are indeed somewhat larger for this part of parameter space, but remain relatively modest. Quantitatively, the $50^{\mathrm{th}}$, $90^{\mathrm{th}}$, and $95^{\mathrm{th}}$ percentiles of $\Delta p$ for young, low-mass clusters are 0.052, 0.406, and 0.544 for the physical network, and 0.018, 0.298, and 0.434 for the photometric network. Based on these statistics, both networks retain good predictive power in the most stochastic regime, with the photometric network providing tighter constraints, consistent with the full validation set. 

\subsection{Completeness as a continuous function}\label{ssec:results_surface}

\begin{figure*}
    \centering
    \includegraphics[width=\textwidth]{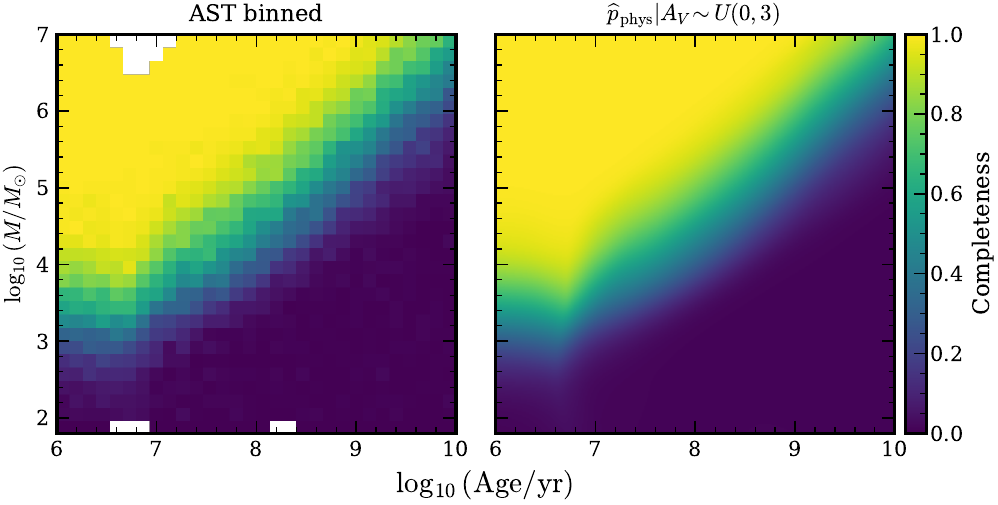}
    \caption{Completeness derived by binning the AST data set (left) versus completeness predicted by the physical neural network (right). The binned prediction is estimated by binning injected clusters in the $(\log M,\log T)$ plane and computing the mean inclusion label per bin, while the NN prediction is evaluated on a fine grid and marginalised over an extinction prior $p(A_V)$ (\autoref{eq:av_marginalisation}). Pixels containing fewer than 10 injected clusters are assigned NaN values and are shown in white in the left panel.}
    \label{fig:compare_bin_nn_libav}
\end{figure*}

\begin{figure}
    \centering
    \includegraphics[width=0.5\textwidth]{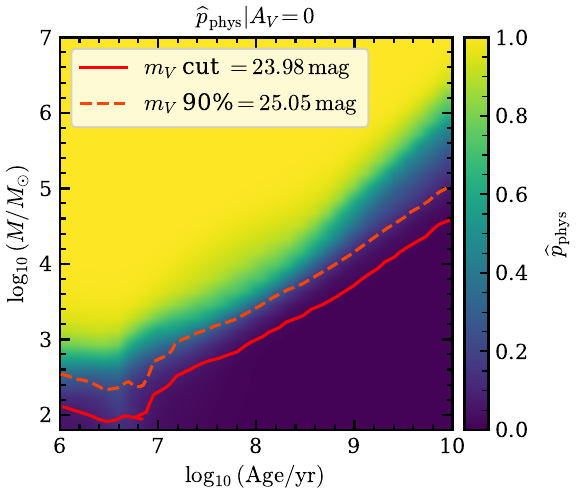}
    \caption{Physical NN-predicted completeness $\widehat{p}_{\rm{phys}}$ surface evaluated at zero $A_V=0$, to consistently compare with literature completeness estimates from \citet{adamo_legacy_2017}. The red solid curve shows the V band magnitude cut mapped into age-mass space, while the orange dashed curve marks the locus of masses and ages corresponding to the $90\%$ completeness limit in V-band magnitude.}
    \label{fig:compare_bin_nn_av0}
\end{figure}

A key advantage of the neural network model is that it provides a smooth, continuous approximation 
\begin{equation}
    \widehat{p}_{\rm phys}(\log M,\log T,A_V)
    \label{eq:nn_phys}
\end{equation}
which gives the predicted inclusion probability for a cluster with physical parameters $(\log M,\log T,A_V)$ to the catalogue selection operator, rather than a lookup table defined on a coarse grid. We visualise this advantage in \autoref{fig:compare_bin_nn_libav}, which compares the empirical completeness calculated directly from the AST training set binned by mass and age to the continuous completeness function predicted by the physical neural network. 

For the grid-based estimate (left panel), we divide the \((\log M,\log T)\) plane into \(30\times30\) uniform bins spanning \(\log (M/\mathrm{M}_\odot) \in [2,7]\) and \(\log (T/\mathrm{yr}) \in [6,10]\), and place clusters from the training (rather than the test) data set into those bins. In each populated bin, the completeness is computed as the arithmetic mean of the binary inclusion label, i.e., the number of detected clusters divided by the total number of injected clusters within that bin. This binned estimator is limited by both the adopted bin resolution and finite sampling noise, which are in tension -- finer bins give more resolution but at the price are large Poisson fluctuations. Our choice of $30\times 30$ bins is our attempt to make a reasonable compromise between these. Also note that, while we are binning in $\log M$ and $\log T$, the results still depend on the assumed underlying distribution of extinction values in the data set used to construct the figure -- a data set more biased to lower $A_V$ would yield higher somewhat higher completeness, one biased to higher $A_V$ would yield lower completeness.

For the NN case (right panel), completeness is evaluated by directly evaluating the learned function \(\widehat p_{\rm phys}(\log M,\log T,A_V)\). 
Since \(\widehat p_{\rm phys}(\log M,\log T,A_V)\) is a three-dimensional function, for the purposes of making the plot we reduce this to the two-dimensional age--mass plane by marginalising over extinction:
\begin{eqnarray}
\lefteqn{
\left.\widehat p_{\rm phys}(\log M,\log T)\right|_{p(A_V)}
=
}
\nonumber \\
& &
\int
\widehat p_{\rm phys}(\log M,\log T,A_V)\,p(A_V)\,\mathrm{d}A_V,
\label{eq:av_marginalisation}
\end{eqnarray}
where $p(A_V)$ is a distribution of extinctions over which to marginalise. For \autoref{fig:compare_bin_nn_libav} we adopt a tophat prior \(A_V \sim U(0,3)\) to match the distribution used in the AST training set. In this case,
\begin{equation}
\widehat p_{\rm phys}(\log M,\log T)
=
\frac{1}{3}
\int_{0}^{3}
\widehat p_{\rm phys}(\log M,\log T,A_V)\,\mathrm{d}A_V.
\label{eq:av_uniform}
\end{equation}
In practice, we evaluate this integral numerically by
sampling the network prediction on a uniformly-spaced regular grid of \(500\times500\times 50\) in \((\log M,\log T, A_V)\) space, then taking the mean over the $A_V$ values.

We see that the NN prediction in \autoref{fig:compare_bin_nn_libav} reproduces the overall structure of the grid-based completeness map while providing a smooth, continuous approximation to the selection boundary. In particular, it captures the gradual transition from high to low completeness across the magnitude-limited regime, while avoiding the fluctuations introduced by finite binning and small-number statistics. An advantage of using neural network approach is that it provides a continuous estimate of completeness across mass--age plane. In the left panel, white pixels indicate NaN values, which we assign to bins containing fewer than 10 injected clusters, highlighting regions that are sparsely sampled by the ASTs. These discontinuities are absent in the right panel because the neural network provides a smooth interpolation across sparsely sampled regions of parameter space.

\subsection{Comparison to literature completeness estimates}

We can also compare the results from our neutral network to earlier attempts to assess completeness in the literature. Prior to this work, the most comprehensive analysis of completeness comes from \citet{adamo_legacy_2017}, who carry out AST tests one photometric band at a time (i.e., not following the full catalogue construction process as we do here), and use the results to estimate their completeness. A primary outcome of this process is Figure~14 of \citeauthor{adamo_legacy_2017}, which provides an estimate of the locus in $(\log M, \log T)$ space where clusters fall below the minimum $V$-band apparent magnitude of 23.98 mag (at the distance of NGC 628) required for including in the LEGUS catalogue, and where the detection probability in $V$-band alone drops below 90\%. We plot these loci on top of our physical NN-predicted completeness in \autoref{fig:compare_bin_nn_av0}. In this case, we evaluate the network at $A_V=0$, i.e., we plot $
\widehat p_{\rm phys}(\log M,\log T,A_V=0)$, to be consistent with \citeauthor{adamo_legacy_2017}, whose estimates are for the zero-extinction case.

We see that the NN-prediction shown in \autoref{fig:compare_bin_nn_av0} is broadly consistent with the literature magnitude-limit curves, but that there are significant differences. In particular, when we reproduce the full LEGUS pipeline, the sensitivity is noticeably worse than one might have estimated based on the single-band testing, particularly for older ages where clusters are redder and thus harder to detect in the bluer bands. One can see hints of this in \citeauthor{adamo_legacy_2017}'s plots of 90\% completeness for different photometric filters (not reproduced in \autoref{fig:compare_bin_nn_av0} to avoid clutter), but the implications of this for overall catalogue completeness cannot be deduced without forward modelling the full pipeline, as we have done here. Other possible contributors to this discrepancy include that previous literature ASTs did not place synthetic clusters according to the galaxy light profile as we do, and did not include the full range of cluster radii that we explore, both of which may lead to overestimates of completeness.


On the other hand, we can see that our NN does capture important, subtle features in the data. One example is the small change in slope around \(\log (T/\mathrm{yr}) \sim 7.2\), which reflects a feature in the stellar population mass-to-light ratio around that produces a local reversal in mapping between a fixed apparent-magnitude threshold and the corresponding cluster mass around that age range. Around this age, the integrated-light properties of the population change rapidly as evolved stars begin to contribute significantly to the broad-band luminosity, leading to a local change in detectability across the LEGUS filter set. The fact that the NN reproduces this feature indicates that it approximates the underlying selection decision boundary accurately enough to retain relatively sharp local transitions, despite the usual spectral bias of neural networks toward smoother structure.


\subsection{Completeness-corrected cluster mass and age distributions}
\label{ssec:results_demographics}

We now use the learned neural-network completeness function to correct the one-dimensional demographic summaries of the observed cluster catalogue, namely the cluster age function (CAF) and cluster mass function (CMF). This has several steps: we first derive marginalised completeness estimates in \autoref{sssec:marginal_completeness}, then define a procedure to make completeness corrections using these marginalised completenesses in \autoref{sssec:completeness_correction_method}. We present the results of applying this procedure to the CAF and CMF in \autoref{sssec:CAF}.


\subsubsection{Marginalisation of the 3D completeness function}
\label{sssec:marginal_completeness}

Our neural network directly predicts the completeness as a function of mass, age, and extinction, $\widehat p_{\rm phys}(\log M,\log T,A_V)$, but to completeness-correct the CMF or CAF, we require the \textit{marginal} completeness as a function of mass or age alone -- that is, to completeness correct, for example, the CMF, we need to know what fraction of the clusters in each mass bin are detectable, but this in turn clearly depends on the intrinsic distribution of cluster ages and extinction in that bin. We obtain these marginalised completenesses under the assumption that the mass, age, and extinction distributions are separable, in which case for the age distribution the effective completeness function marginalised over the cluster mass and extinction is
\begin{eqnarray}
\lefteqn{
\widehat p_{\rm corr}(\log T)
=
}
\nonumber \\
& 
\iint
\widehat p_{\rm phys}(\log M,\log T,A_V)\,
p(\log M)\,p(A_V)\,
\mathrm d\log M\,\mathrm dA_V,
\label{eq:pcorr_age_marg}
\end{eqnarray}
where $p(\log M)$ and $p(A_V)$ are the \textit{intrinsic} mass and extinction distributions over the cluster population. The equivalent expression required correct the observed CMF, $\widehat p_{\rm corr}(\log M)$, is defined analogously by marginalising over $p(A_V)$ and the intrinsic age distribution $p(\log T)$. To evaluate these integrals, we therefore must make estimates for the intrinsic mass, age, and extinction distributions.\footnote{Careful readers may worry that this is circular, since we are assuming intrinsic distributions in order to make completeness corrections. However, there is no circularity, because we use assumed intrinsic mass distribution \textit{only} to completeness-correct the age distribution, and vice-versa.} 

To estimate the intrinsic extinction distribution, we define a high-completeness region in the \((\log M,\log T)\) plane using our trained physical neural network evaluated at fixed extinction \(A_V=3\), corresponding to the upper limit of the extinction range considered here and therefore providing a conservative selection boundary where we can be confident that even high-extinction clusters are visible. Specifically, we evaluate \(\widehat p_{\rm phys}(\log M,\log T, A_V=3)\) on a fine grid in \((\log M,\log T)\), construct a contour corresponding to \(\widehat p_{\rm phys}(\log M,\log T, A_V=3) = 0.5\), and retain only those observed clusters whose masses and ages fall within that contour. We then estimate \(p(A_V)\) from the $A_V$ values of the retained clusters by creating a histogram in $A_V$. Because there are only 14 clusters within our high completeness region, we use a very coarse binning, with four uniformly-spaced bins from $A_V = 0 - 3$; we then take $p(A_V)$ to be a piecewise-constant function corresponding to this histogram, properly normalised to $\int p(A_V) \, \mathrm{d}A_V = 1$.

We consider two choices each for $p(\log M)$ and $p(\log T)$, both based on earlier analyses of the cluster population of NGC~628: one taken from the best-fitting models of \citet{2024Tang}, and one from the fits provided by \citet{adamo_legacy_2017}. For the mass distribution, the former gives a Schechter function form $p(\log M) \propto M^{\alpha_M+1} \exp(-M/M_\mathrm{break})$ with \(\alpha_M=-2.16\) and \(\log (M_{\rm break}/\mathrm{M}_\odot)=5.61\), while the latter gives pure powerlaw form $p(\log M)\propto M^{\alpha_M+1}$ with $\alpha_M = -2.09$. For $p(\log T)$, \citeauthor{2024Tang}'s best-fit is a broken powerlaw form $p(\log T) \propto T$ for $\log (T/\mathrm{yr}) < 8.22$ and $p(\log T)\propto T^{\alpha_T + 1}$ with $\alpha_T = -1.41$ for older ages, while \citeauthor{adamo_legacy_2017} adopt a single powerlaw form $p(\log T)\propto T^{\alpha_T+1}$ with $\alpha_T = -0.8$ over the age range $\log(T/\mathrm{yr})\in [6.0,8.3]$.

Armed with these estimates for the true distribution, we compute our marginalised completeness functions $\widehat p_{\rm corr}(\log T)$ and $\widehat p_{\rm corr}(\log M)$ by evaluating the integral \autoref{eq:pcorr_age_marg}, and the analogous expression for mass, using a discrete summation approximation evaluated on a uniform grid of 500 points in $\log M$ or $\log T$ by our 4 bins in $A_V$.

\subsubsection{Bin-level completeness correction}
\label{sssec:completeness_correction_method}

Once the one-dimensional effective completeness functions have been constructed, we evaluate them for each observed cluster according to the quantity being corrected. For this exercise, we adopt the individual cluster masses and ages inferred by \citet{adamo_legacy_2017}, and for the \(i^{\rm th}\) cluster we use \(\widehat p_{\rm corr}(\log T_i)\) when correcting the CAF and \(\widehat p_{\rm corr}(\log M_i)\) when correcting the CMF. Denoting these object-level effective completeness values generically by \(\widehat p_{{\rm corr},i}\), we compute bin-level corrected counts using an inverse-completeness correction,
\begin{equation}
N_{\rm corr,bin}
=
\frac{N_{\rm raw,bin}}{\overline p_{\rm corr,bin}},
\qquad
\overline p_{\rm corr,bin}
=
\frac{1}{N_{\rm raw,bin}}
\sum_{i\in b}\widehat p_{{\rm corr},i}.
\label{eq:bin_level_correction}
\end{equation}
Here, \(N_{\rm raw,bin}\) is the observed number of clusters in a given age or mass bin, \(N_{\rm corr,bin}\) is the corresponding completeness-corrected count, and \(\overline p_{\rm corr,bin}\) is the mean effective completeness of the clusters in that bin. To avoid numerical instability caused by very small completeness values, we impose a lower clipping threshold on \(\widehat p_{{\rm corr},i}\) before applying the correction. For the CAF we clip the values of $\widehat p_{{\rm corr},i}$ to a minimum of \(10^{-10}\) for individual objects, and mask bins for which the mean completeness \(\overline p_{\rm corr,bin}<10^{-3}\); for the CMF correction, the per-object completeness is clipped at \(10^{-3}\). This clipping affects only a small fraction of objects ($< 3\%$), but prevents individual sources from dominating the corrected counts through excessively large inverse-completeness weights.

To quantify the uncertainty in the completeness correction within each bin, we compute the $16^{\rm th}$ and $84^{\rm th}$ percentiles of the set of object-level completeness values, $\{\widehat p_{\mathrm{corr},i}\}_{i\in j}$, in each bin $j$, denoted by $p_{16,j}$ and $p_{84,j}$. We then insert these values into \autoref{eq:bin_level_correction} in place of the mean, $\overline p_{\rm corr,bin}$, to obtain alternative upper and lower values for each bin,
\begin{equation}
N_{{\rm lo},j}
=
\frac{N_{{\rm raw},j}}{p_{84,j}},
\qquad
N_{{\rm hi},j}
=
\frac{N_{{\rm raw},j}}{p_{16,j}}.
\end{equation}
Thus $N_\mathrm{lo}$ and $N_\mathrm{hi}$ for each bin amount to the result we would have obtained if we were to set the completeness corrections for all objects in that bin to the 84th or 16th percentile values for objects in that bin. This is not a formal error estimate, but gives a rough sense of the level of uncertainty induced by our completeness correction.

\subsubsection{Completeness-corrected cluster age and mass functions}
\label{sssec:CAF}

\begin{figure}
    \centering
    \includegraphics[width=0.5\textwidth]{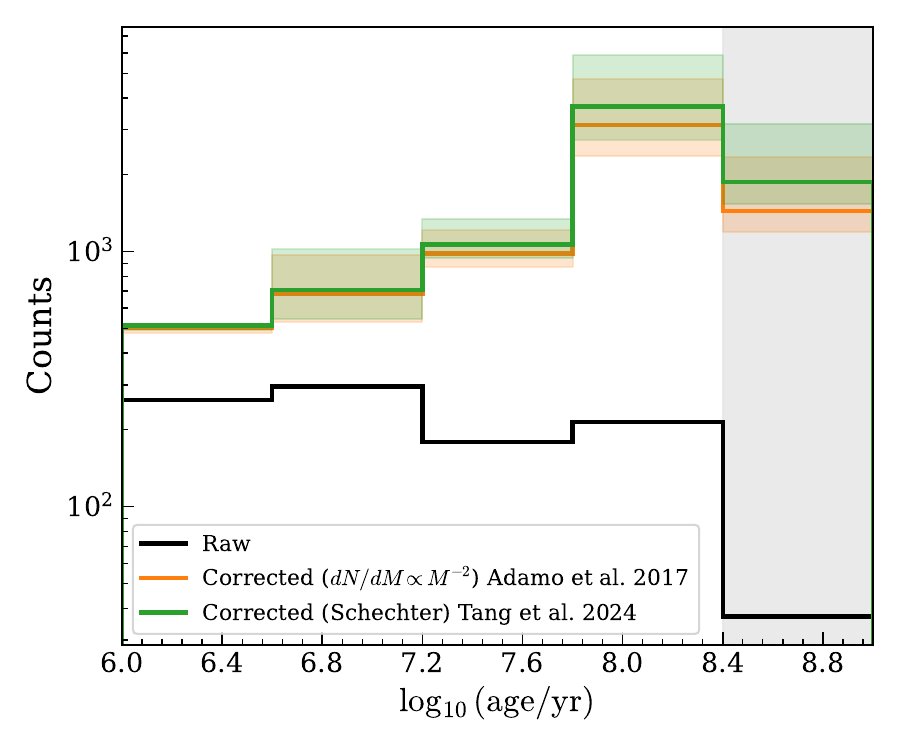}
    \caption{Completeness-corrected CAF. The black line shows that raw (uncorrected) age histogram, while orange and green lines show completeness-corrected CAFs assuming the CMFs from \citet{adamo_legacy_2017} and \citet{2024Tang}, respectively. Shaded envelopes show the confidence intervals between the $16\%$ and $84\%$ percentiles of correction factors within each bin. The vertical shaded band shows ages $>200$ Myr, beyond the maximum age generally used for demographic analysis even for massive clusters \citep[e.g.,][]{adamo_legacy_2017}}
    \label{fig:CAF}
\end{figure}

\autoref{fig:CAF} shows the raw and completeness-corrected CAFs computed for both our assumed intrinsic CMFs; the completeness-corrected estimates include uncertainties obtained as described above. We see that both of the possible CMF models lead to similar results within the uncertainties. As expected, the completeness correction increases systematically toward older ages, where clusters are dimmer and thus more likely to be missed. Consequently while the raw age distribution is relatively flat with age, corresponding to strong evidence for cluster disruption (since flat in log age implies a distribution in linear age $dN/dT \propto T^{-1}$), the completeness-corrected distribution rises mildly with age, with a slope $\sim 0.3$, corresponding to $dN/dT \sim T^{-0.7}$. This indicates somewhat milder disruption than would be inferred for an uncorrected sample.




\begin{figure}
    \centering
    \includegraphics[width=0.5\textwidth]{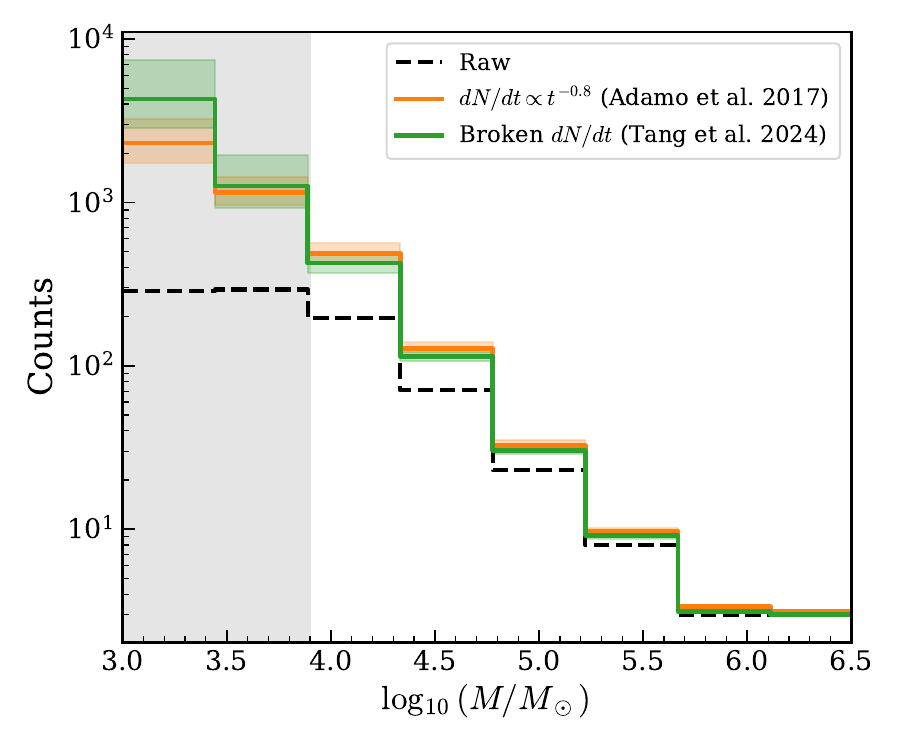}
    \caption{
    Same as \autoref{fig:CAF}, but now showing the completeness-corrected CMF. The shaded grey band shows masses $<5000$ M$_\odot$, the traditional mass limit used in earlier analyses \citep[e.g.,][]{adamo_legacy_2017}.}
    \label{fig:CMF}
\end{figure}

\autoref{fig:CMF} presents analogous results for the completeness-corrected CMF, again comparing the raw (uncorrected) and completeness-corrected CMFs, with the latter obtained under two alternative assumptions for the age-function. From \autoref{fig:CMF} it is clear that the completeness correction is strongly mass-dependent. At low masses, the correction factor reaches $\sim 10$, reflecting the rapidly declining inclusion probability toward the detection boundary. The completeness correction removes the flattening seen at low masses in the raw counts, and reveals a distribution consistent with an unbroken powerlaw down to at least $10^3$ M$_\odot$, beyond which completeness correction becomes nearly impossible because essentially no clusters are observed. Above $M \sim 10^{5},{\rm M}_{\odot}$, the inferred completeness approaches unity and the correction becomes negligible. In this regime the cluster sample is effectively complete, independent of the assumed age distribution $p(\log T)$ used in the marginalisation. As with CAF, we see relatively little difference between results derived using the two possible assumed functional forms of the CAF.

The vertical grey shaded region in both \autoref{fig:CAF} and \autoref{fig:CMF} mark the limits commonly adopted in earlier analysis of cluster demographics, $M > 5\times10^{3}\,{\rm M}_{\odot}$ and $T < 200$ Myr \citep[e.g.,][]{adamo_legacy_2017}. Earlier authors limited their demographic analysis to clusters satisfying \textit{both} of these conditions, on the grounds that either older or less massive regions of the $(\log M, \log T)$ plane would be highly incomplete. 
Our results demonstrate that neural-network-based completeness modelling removes this limitation, and provides a robust and flexible alternative to traditional binning approaches, enabling accurate corrections into previously-inaccessible parts of parameter space.

\subsection{The effect of metallicity}
\label{ssec:metallicity}

All the results we have presented thus far are derived using the Solar-metallicity SLUG library appropriate for NGC~628. It is therefore interesting to consider how metallicity affects completeness estimates, and how one might relax the assumption of uniform metallicity in future work. First, note that in the photometric network metallicity does not appear explicitly because completeness is conditioned directly on observed magnitudes and colours. Changing the metallicity would change the distribution of the training set in photometric space (since cluster colours depend on metallicity), but this seems unlikely to have any significant effect on the inferred completeness at a fixed point in photometric space. The effects are likely to be more prominent in physical space, since metallicity affects the mapping from $(M,T,A_V)$ to photometry, and can therefore change the inferred completeness. For example, adopting a Sub-solar stellar-population library, such as $Z = 0.004$, would generally reduce metal-line blanketing and shift young populations blueward, increasing their UV/blue luminosities at fixed mass, age, and extinction \citep[e.g.,][]{Leitherer2001_UVMetalPoor,Choi2016_MISTI}; SLUG propagates these stellar-population ingredients into stochastic cluster photometry \citep{2012SLUGI,2015SLUGIII}.

The effect of such a shift on completeness in any given survey will depend on the relative depths of the observations in different bands, and on how the catalogue construction pipeline handles situations where a cluster is detected in some bands but not in others. In the case of LEGUS, which requires a detection in the UV or U bands for catalogue inclusion, and where there are a number of cluster candidates detected in the redder bands that are excluded from the final catalogue because they are missed in both the bluer bands, the blueward shift induced by lower metallicity would likely increase completeness at fixed $(M,T,A_V)$ compared to the Solar metallicity case. However, we emphasise that this is a statement particular to LEGUS, and may not apply to other surveys with different sensitivies and catalogue construction strategies.

In principle one could also extend our method by treating metallicity $Z$, or more broadly stellar chemical composition, as an additional physical parameter in the neural network. This would require drawing from a range of $Z$ rather than a single one when constructing the SLUG library, and then using that library to learn $\widehat p_{\rm phys}(\log M,\log T,A_V,\log Z)$ rather than just $\widehat p_{\rm phys}(\log M,\log T,A_V)$ as in our current implementation. In practice, our ability to pursue the strategy is limited by the availability of the underlying models on which SLUG relies. Treating $\log Z$ as a continuously-sampled variable like $\log M$, $\log T$, and $A_V$ would require the ability to generate stellar population models at arbitrary $\log Z$, which in turn depends on the ability to generate such models on the fly or credibly interpolate between models computed at fixed values of $\log Z$. While grids of stellar evolution model tracks that are densely-enough sampled in metallicity to allow credible interpolation are now available (e.g., MIST provides samples at spacings of 0.25 in $\log Z$; \citealt{Dotter2016_MIST0, Choi2016_MISTI}), the situation is much worse for stellar atmosphere models, particularly for O and Wolf-Rayet stars that do not have hydrostatic atmospheres and for which UV fluxes therefore depend sensitively on the hydrodynamic structure of the wind and the associated line blanketing effects those induce \citep[e.g.,][]{Leitherer1999_Starburst99,Smith2002_IonizingFluxes,Conroy2013_SPSReview}. There are at present no publicly available grids of atmosphere models for such stars with dense enough sampling for sensible interpolation, and thus we currently lack the ability to generate training sets with continuous metallicity distributions suitable for training neural networks in which metallicity is treated as a continuous, free parameter.

\section{Conclusions}
We have developed and validated a general framework, the Cluster Completeness Correction Calculator (C-4), for modelling completeness in extragalactic star cluster catalogues using neural networks. Our approach combines artificial star cluster tests (AST) that replicate the full cluster detection and and filtering pipeline, thereby reproducing the complete selection pathway required for a cluster to enter the final catalogue. 
We then use the results of these tests as input to a multi-layer perceptron neutral network, which learns the inclusion probability directly as a function of intrinsic cluster properties -- either physical or photometric. We show that the resulting networks yield highly accurate predictions for the results of completeness tests for clusters over the full range of physical and photometric properties spanned by real star clusters.

Compared to traditional binning approaches and parametric completeness curves, neural networks offer several key advantages. First, they capture non-linear interactions between mass, age, extinction, and photometric observables without requiring explicit parametric assumptions. Second, they provide a continuous and differentiable approximation to the selection function, avoiding discretisation artefacts introduced by binning. Third, they naturally incorporate the stochastic processes arising from IMF sampling. The resulting completeness function is therefore a smooth probabilistic operator rather than a discretised or threshold-based approximation.

The primary outcome of this pipeline is a three-dimensional completeness function that is continuous, marginalisable, and differentiable in both physical -- mass, age, and extinction space -- and photometric space. In physical space, we show that completeness exhibits a highly structured and multivariate geometry in the $(\log M, \log T, A_V)$ plane, particularly near the detection boundary, which cannot be captured by simple binning schemes. The neural network model instead provides a stable and physically consistent representation suitable for forward modelling.

In this paper, we provide a first pilot application of C-4 to the LEGUS catalogue of NGC~628 \citep{2015LEGUS, 2015ApJ...815...93G}, a system with a well-characterised cluster population. For this system we show that the learned completeness function qualitatively agrees with earlier results for completeness while yielding a smoother and more stable correction near the detection boundary, and a more accurate treatment of the interacting effects of detection thresholds across multiple photometric bands that were ignored in earlier analyses. When applied to the cluster mass and age functions in NGC~628, the corrected CMF remains consistent with a power-law form down to masses roughly an order of magnitude below the limits of previous studies, while the CMF becomes somewhat more positive in slope, indicating reduced evidence for cluster disruption at ages roughly an order of magnitude beyond the range probed by earlier studies.


In future work, we will extend this framework to the full LEGUS sample to perform a homogeneous and selection-consistent study of cluster populations across diverse galactic environments. This establishes neural-network-based completeness modelling as a robust alternative to traditional binning approaches for analysing stellar cluster populations. More broadly, this work reframes completeness estimation as a learnable, high-dimensional selection problem rather than a binning exercise, and provides a pathway toward fully forward-modelled analyses of star cluster populations in the JWST era.
\label{sec:conclusions}

\section*{Acknowledgements}
We thank the anonymous referee for their valuable comments and suggestions. JT and KG are supported by the Australian Research Council (ARC) through the Discovery Early Career Researcher Award (DECRA) Fellowship (project number DE220100766) funded by the Australian Government. MRK is supported by the ARC through Laureate Fellowship FL220100020. This research was supported by the National Computational Infrastructure (NCI), which is supported by the Australian Government, through the National Computational Merit Allocation Scheme and the ANU Merit Allocation Scheme (award jh2). Based on observations made with the NASA/ESA Hubble Space Telescope, obtained at the Space Telescope Science Institute, which is operated by the Association of Universities for Research in Astronomy, Inc., under NASA contract NAS 5-26555. These observations are associated with program \#13364.

\section*{Data Availability}
The cluster catalogue data underlying this paper are publicly available online at MAST. The software suite, tools, and dependencies required to perform the full completeness pipeline, as well as the weights and biases of the trained neural networks, are publicly accessible on \href{https://github.com/JianlingTang/completeness_NN_pipeline}{GitHub}. In addition, we package the neural networks as Python-based calculators. The package is available on PyPI and can be installed via \texttt{pip install cluster-completeness-pipeline}. 
A detailed description of the inputs, outputs, and function arguments is provided in the package documentation.



\bibliographystyle{mnras}
\bibliography{example} 




\appendix

\section{Construction of white light images for source extraction}\label{app:white_light}
Here we describe the procedure used to construct the white-light image from the five-band LEGUS \textit{HST} imaging of NGC~628C. We begin with the science images in the F275W, F336W, F435W, F555W, and F814W filters. For each filter image $f_\lambda$, we estimate the median and dispersion of the background from the pixel-value distribution, adopting a fixed detector read noise of $3.1\,e^-$. The sky-subtracted image is then obtained by subtracting the median background level and clipping negative pixel values to zero,
\[
f_\lambda' = \max(f_\lambda - \mathrm{median}(f_\lambda), 0),
\]
where the background-subtracted images are denoted by $f_\lambda'$.

We then combine the five sky-subtracted images into a single white-light image using a weighted quadrature sum, i.e., we write the white light intensity $I$ in a given pixel as
\begin{equation}
    I = \frac{\left[\sum_\lambda (f'_\lambda/s_\lambda)^2 \right]^{1/2}}{\left[\sum_\lambda (1/s_\lambda)^{2}\right]^{1/2}},
\end{equation}
where $f'_\lambda$ represents the sky-subtracted pixel values in the $\lambda = 1-5$ available filters and $s_\lambda$ is a set of scaling factors used to weight the filter images. Following the original \citet{2015LEGUS} LEGUS pipeline, we apply two different sets of $s_\lambda$ values, which we list in \autoref{tab:white_light_generation}. These two sets of scale factors are chosen to maximise signal to noise ratios for stellar populations dominated by stars at two distinct phases of stellar evolution: bright 21 Myr main-sequence stars and RGB stars. We refer readers to \citeauthor{2015LEGUS} for discussion of the motivation for choosing these phases and the method used to optimise the scale factors for them; we do not discuss them here, since for our purposes the only important thing is that we match the values used in original observational pipeline. Once we have constructed white-light images using the two different sets of scale factors, we simply take the arithmetic mean of the two as our final white-light image that is passed to the next step of the detection pipeline.

\begin{table}
\centering
\caption{Scale factors $s_\lambda$ used to construct the white-light image from multi-band \textit{HST} science images. The two columns give the scale factors determined to optimise detection of 21 Myr-old main sequence and RGB (red-giant branch) populations, respectively.}
\label{tab:white_light_generation}
\begin{tabular}{lcc}
\hline
Filter & 21 Myr & RGB \\
\hline
F275W & 55.8 & 0.5 \\
F336W & 45.3 & 0.8 \\
F435W & 44.2 & 3.1 \\
F555W & 65.7 & 11.2 \\
F814W & 29.3 & 13.7 \\
\hline
\end{tabular}
\end{table}

\section{Construction of the synthetic cluster library using SLUG}

As discussed in \autoref{ssec:generation_of_ASTs}, we generate our artificial cluster library using the
\textsc{slug} stochastic stellar population synthesis code \citep{2012SLUGI, 2015SLUGIII}. In addition to the star cluster mass function, age distribution, and extinction distribution already discussed in \autoref{ssec:generation_of_ASTs}, \text{slug} requires that we set a number of additional parameters, which we document here. Except as noted below, the numerical implementations of these choices follow the methods outlined in \citet{2015SLUGIII}. These are:

\textbf{Extinction curve.} We adopt \text{slug}'s Milky Way extinction curve \citep{1984MWExt, 1999MWExt}.

\textbf{Initial mass function (IMF).} We draw stars from a \citet{2005ChabrierIMF} IMF from $0.08$ to $120\,M_{\odot}$. This IMF consists of lognormal with mean $0.2\,M_{\odot}$ and dispersion $\sigma=0.55$ dex at masses $<1$ M$_\odot$ and a power law with slope $-2.35$ at higher masses. 

\textbf{Stellar tracks.} We use MIST v1.0 stellar tracks for stars rotating at $40\%$ of breakup velocity at birth \citep{Dotter2016_MIST0, Choi2016_MISTI}.

\textbf{Metallicity.} We adopt solar metallicity ($Z=0.02$) for the stellar tracks described above.

\textbf{Stellar atmospheres.} We adopt \textsc{slug}'s
``starburst99'' model for stellar atmospheres at Solar metallicity, which uses the same set of libraries and rules for determining which libraries are used for which star as the \textsc{starburst99} code \citep{Leitherer1999_Starburst99}.

\textbf{Nebular emission.} We include nebular emission, and set the fraction of the ionizing photon flux that is reprocessed into nebular emission within the observed aperture to $\phi=0.5$ and the log ionisation parameter to $\log\mathcal{U} = -3.$

Our treatment of nebular emission here differs from that in the original \citeauthor{2015SLUGIII} \textsc{slug} method paper as a result of a recent upgrade to the code. This change is documented in the \textsc{slug} user manual\footnote{\url{https://slug2.readthedocs.io/en/latest/}} but has not previously been described in the published literature, so we summarise it briefly here.

To compute nebular emission, \textsc{slug} relies on a set grid of model H~\textsc{ii} regions computed using \textsc{cloudy} \citep{Chatzikos23a}. The models, which are computed for each IMF and metallicity available in \textsc{slug} (though we use only Solar metallicity models in the present-work), span a 2D grid in the space of stellar population age, from 0 to 10 Myr in steps of 0.2 Myr, and volume-averaged ionisation parameter $\mathcal{U}$ at values of $\log\mathcal{U} = -3$, $-2.5$, and $-2$ (with $\log\mathcal{U}=-3$ used for all the models in this paper); the first of these parameters determines the shape of the ionising spectrum, and the second the ratio of the ionising luminosity to the gas density.\footnote{Unlike in the previous treatment of nebular emission in \textsc{slug}, we do not treat the gas density itself as a separate variable. As long as the density is below the critical density of lines bright enough to contribute significantly to the broadband colours ($\gtrsim 10^5$ cm$^{-3}$), we find that variations in the density at fixed $\log\mathcal{U}$ have very little effect on the broadband emission.} See \citeauthor{2015SLUGIII} for an explanation of how these parameters translate to the inputs to the \textsc{cloudy} calculation. For each model in the grid, \textsc{cloudy} is run to convergence, and then the computed nebular spectrum is divided into continuum and line emission components. We separately record the line luminosity per ionising photon for the 100 brightest lines and the full spectral shape of the continuum, again normalised to the ionising photon luminosity.

To generate predictions for the nebular emission from this grid, for any star cluster with an age $< 10$ Myr, we compute the ionising luminosity in the stellar spectrum, multiply by $\phi$ to account for loss of ionising photons to dust absorption or escape, interpolate on the grid of stellar population age and ionisation parameter to obtain the line and continuum luminosity per ionising photon for the cluster, and finally multiply by the ionising luminosity to obtain the nebular contribution to the spectrum. We apply the calculated extinction $A_V$ to both the stellar and nebular emission, with a factor of $2.1$ larger extinction applied to the nebular light compared to the stellar light (see \citealt{2019SLUGIV} for details), and add the extincted stellar and nebular spectra to obtain the final observed spectrum. 

Finally, we caution that, as with the original implementation of fast nebular emission in \textsc{slug}, this treatment does not properly capture the effects of stochastic variation in the shape of the stellar spectrum arising from IMF sampling. This effect is important for emission in high-ionisation lines, but is relatively unimportant for the broadband colours of interest in the present work -- see \citet{Arora21a} for a detailed discussion.

\label{app:slug_lib}

\section{Sensitivity to mass-radius relation}

The AST procedure requires an assumption about the mass–radius relation when assigning effective radii to synthetic clusters. To assess whether this structural dependence significantly affects the inferred selection operator, we repeat the full AST and neural-network training procedure under the two alternative prescriptions defined in \autoref{ssec:generation_of_ASTs}.

\begin{figure*}
    \centering
    \includegraphics[width=\textwidth]{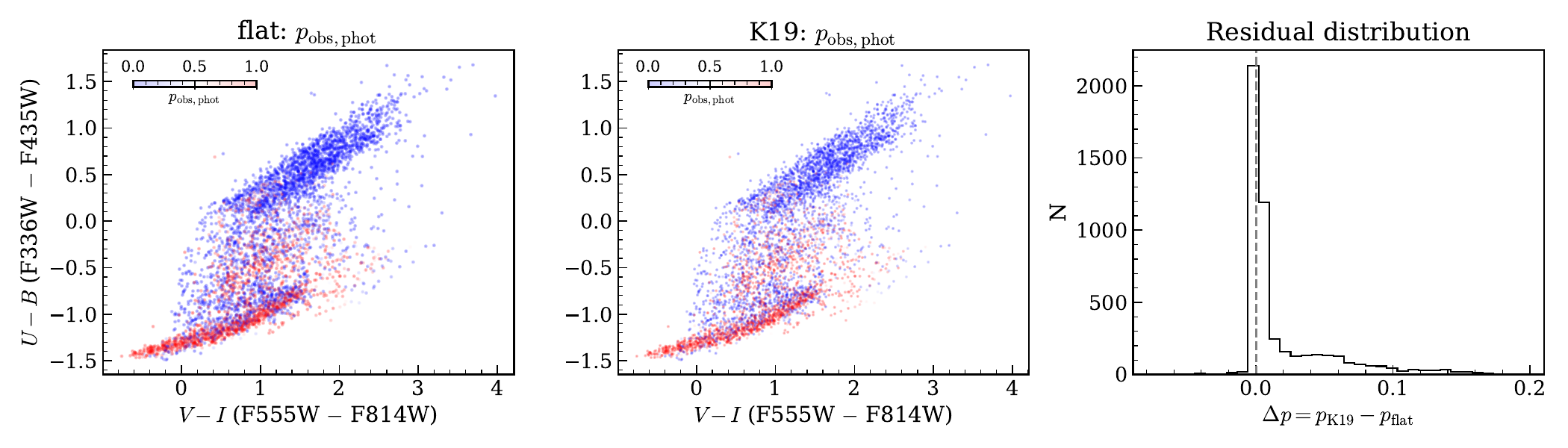}
    \caption{Predicted completeness evaluated for cluster sample that has not been used in training under two alternative structural prescriptions for synthetic-cluster generation. Left and middle panels show completeness in the (V-I, U-B) plane for the constant-radius and $r_{\rm eff}\propto M^{1/3})$ assumptions, respectively. The right panel shows the distribution of residuals of predictions using two prescriptions, demonstrating minimal dependence of the inferred selection operator on the adopted mass–radius relation.}
    \label{fig:color_compare}
\end{figure*}

To see how strongly the resulting networks differ, we apply the photometric networks to our reserved sample of 5000 test artificial clusters (which have not been used to train or validate the networks). In the left and middle panels of \autoref{fig:color_compare} we compare the predicted completeness for these clusters as a function of their location in the $(V-I)$ versus $(U-B)$ colour plane. The left panel shows predictions obtained using the flat mass-radius prescription, while the middle panel shows results from our $r_{\rm eff} \propto M^{1/3}$ relation; we denote this the K19 model, since the relation is inspired by the data collected in the review by \citet{2019ARA&A..57..227K}. The right panel shows the distribution of differences in predicted completeness, $\Delta p = p_\mathrm{K19} - p_{\rm flat}$. We see that the predicted $p_\mathrm{obs}$ values are qualitatively similar, and that the distribution of residuals is strongly peaked near zero: the central two bins around $\Delta p = 0$ contain roughly 2/3 of the total sample. The remainder fall mostly into a small tail of positive $\Delta p$ values that peaks around $\Delta p \sim 0.05$, indicating marginally higher detection probabilities for the K19 mass-radius relation. This occurs because the flat prescription yields a larger number of extended low-mass clusters that are harder to detect, whereas the K19 prescription produces fewer such objects. However, the difference overall is quite modest.


\label{app:mass-radius}

\section{Neural network training details}
\label{app:training}

Here we provide further details on how we train our neural networks, including hyperparameter optimisation (\aref{app:hyperparameters}) and loss rates (\aref{app:training_details}).

\subsection{Selection of neural network hyperparameters}
\label{app:hyperparameters}

A key part of training machine-learning models is choosing suitable hyperparameters. For our chosen network architecture, we optimised two: the maximum learning rate $\eta_\mathrm{max}$ and weight decay coefficient $\lambda$. We employ a grid search to identify values of these hyperparameters that minimise the loss. In our grid, we test maximum learning rates $\eta_\mathrm{max} = [5\times10^{-3}, 1\times10^{-2}, 5\times10^{-2}, 1\times10^{-1}]$,  weight decays $\lambda = [1\times10^{-3}, 5\times10^{-3}, 1\times10^{-2}, 3\times10^{-2}, 5\times10^{-2}, 1\times10^{-1}]$. For each combination of these parameters, we train for 100 epochs using a batch size of 4096, and compute the final validation loss $\mathcal{L}_\mathrm{val}$ achieved at the end of this process.

\begin{figure}
    \centering
    \includegraphics[width=0.5\textwidth]{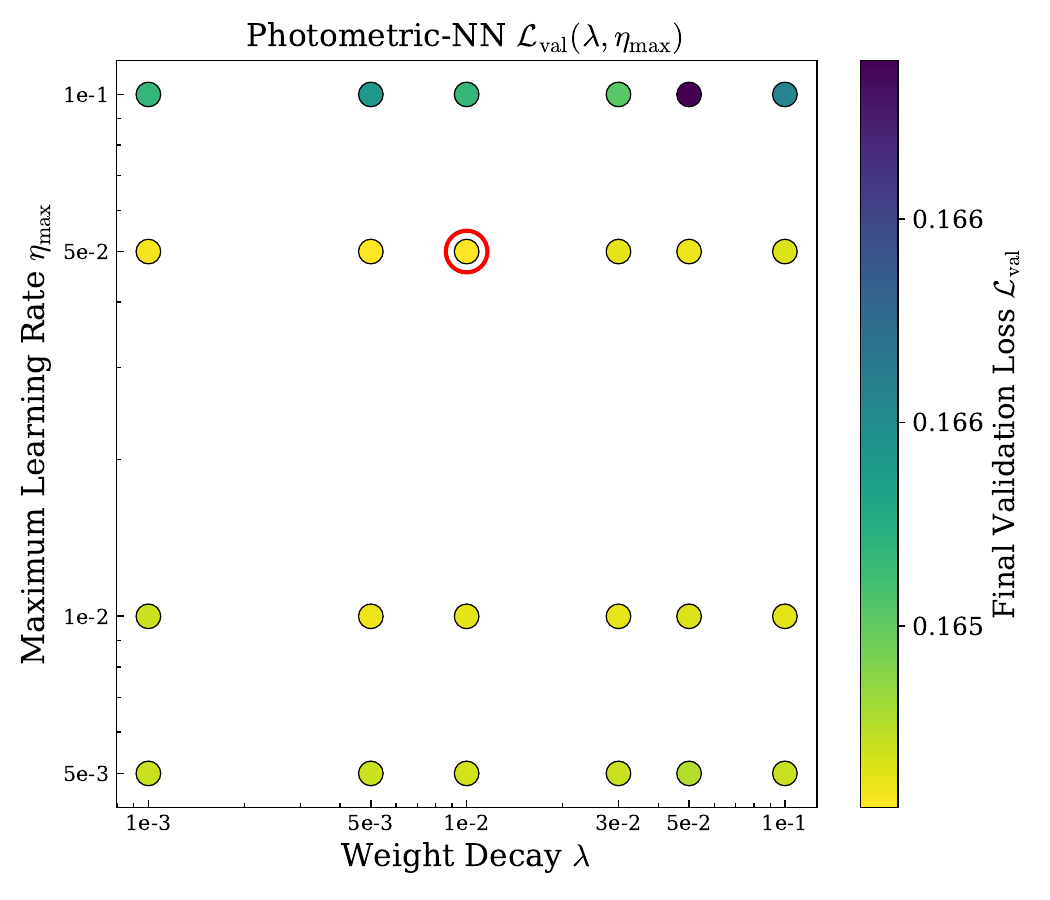}
    \caption{Scatter plot of the final validation loss for the photometric neural network evaluated for various combinations of weight decay and maximum learning rate. The combination set of weight decay and maximum learning rate values that yield the minimum final validation loss is circled in red.}
    \label{fig:hyperparameters_phot}
\end{figure}

\begin{figure}
    \centering
    \includegraphics[width=0.5\textwidth]{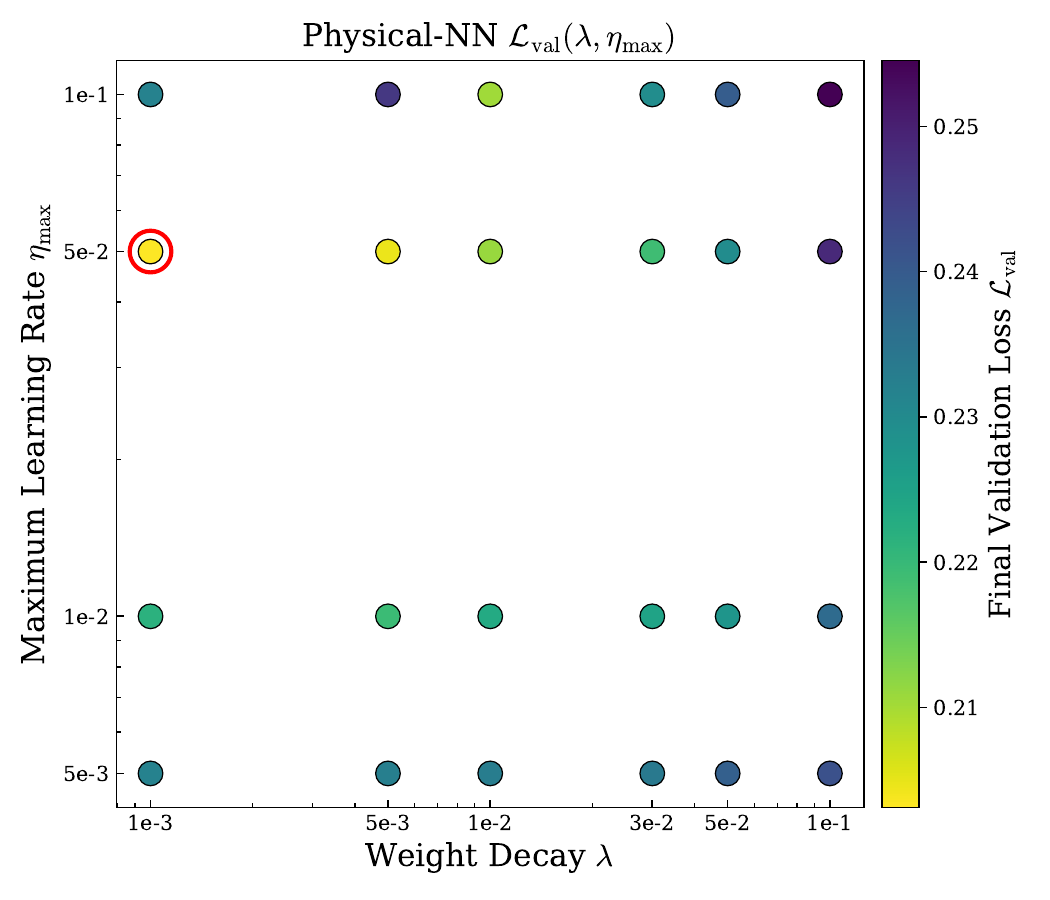}
    \caption{Same as \autoref{fig:hyperparameters_phot} but for the physical neural network.}
    \label{fig:hyperparameters_phys}
\end{figure}

We plot $\mathcal{L}_\mathrm{val}$ as a function of $\eta_\mathrm{max}$ and $\lambda$ in \autoref{fig:hyperparameters_phot} and \autoref{fig:hyperparameters_phys} for the photometric and physical networks, respectively. We select the parameter combination that produces the minimum validation loss (circled in red in the figure) for each network, and use this set of parameters for the remainder of this work. These best parameters are $(\eta_\mathrm{max}, \lambda) = (5\times 10^{-2}, 1\times 10^{-3})$ for the photometric network and $(1\times 10^{-2}, 5\times 10^{-2})$ for the physical network.

\subsection{Loss rates}
\label{app:training_details}

\begin{figure*}
    \centering
    \includegraphics[width=\textwidth]{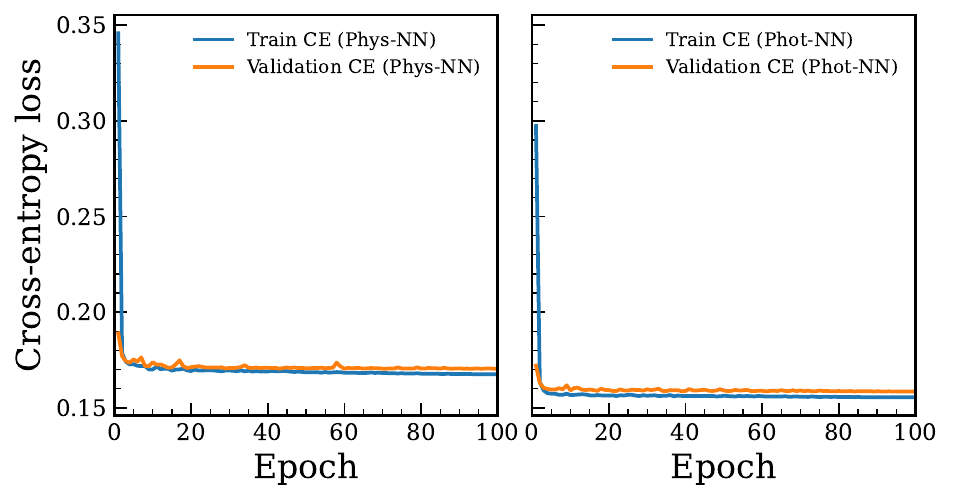}
    \caption{Cross-entropy loss as a function of training epoch for physical (left panel) and photometric (right panel) neural networks, showing both training (orange) and validation (blue) sets.}
    \label{fig:phys_phot_best_ce}
\end{figure*}

To ensure that our network is learning well and is not over-fitting, in \autoref{fig:phys_phot_best_ce} we present the plots of cross-entropy loss versus epochs in training for both the photometric and physical networks. Both neural networks reach a plateau in cross-entropy loss after approximately 100 epochs, showing no evidence of overfitting.

\begin{figure}
    \centering
    \includegraphics[width=0.5\textwidth]{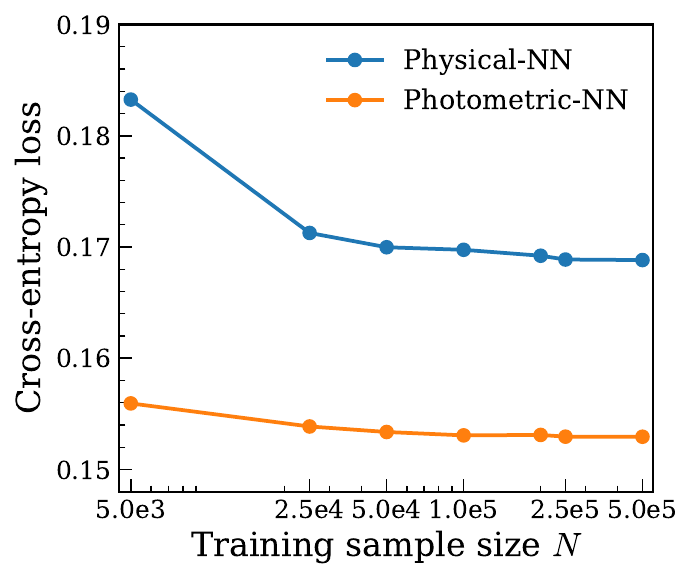}
    \caption{Validation cross-entropy loss after 100 epochs versus training sample size for the physical (blue solid line) and photometric (orange solid line) neural networks.}
    \label{fig:ValCE_sample_size}
\end{figure}

To ensure that our training sample size is large enough, we repeat the training procedure with different sample sizes from $5\times10^3$ to $5\times 10^5$ -- for samples smaller than the fiducial $2.5\times 10^5$ used in the main text we select subsets of the training set at random, while for the test with $5\times 10^5$ clusters we generate we new training set equal in size to the one used in the main text, using the same method, and train on the combination of the two. We show the validation cross-entropy loss after 100 epochs as a function of training sample size in \autoref{fig:ValCE_sample_size}. The plot shows that performance improves rapidly at small $N$ and begins to plateau at $\approx 5\times10^{4}$. Beyond this point, doubling the training set yields marginal improvement in validation loss $\Delta\mathrm{CE} \lesssim 0.001$, while substantially increasing computational cost. Although the validation loss begins to plateau at $N \approx 5\times10^{4}$, the curves continue to decrease slightly up to $N=2.5\times10^{5}$, while the improvement beyond this point is minimal. We therefore adopt a sample size of $N=2.5\times10^{5}$ as a practical balance between convergence efficiency and predictive performance. This choice is broadly consistent with sample sizes reported in previous work \citep[e.g.,][]{2024AJ....168...38H}. 

\bsp	
\label{lastpage}
\end{document}